  \documentclass[structabstract]{aa}  
  \usepackage{graphicx}
  \usepackage{txfonts}
  \usepackage{amssymb}
  \usepackage{natbib}
  \usepackage{lscape}

  \bibpunct{(}{)}{;}{a}{,}{,}
  \usepackage{times}

  \def\min{${}^{\prime}$}
  \def\sec{${}^{\prime\prime}$}
  \usepackage{longtable}

\begin{document}
    \title{Star formation in the outer Galaxy: membership and fundamental
  parameters of the young open cluster NGC\,1893 }

    \author{L. Prisinzano\inst{1}
	  \and
	  J. Sanz-Forcada\inst{2}
	  \and
	  G. Micela\inst{1}
	  \and
	  M. Caramazza\inst{1}
	  \and
	  M. G. Guarcello\inst{1}
	  \and
	  S. Sciortino\inst{1}
	  \and
	  L. Testi\inst{3}
	  }

    \institute{INAF - Osservatorio Astronomico di Palermo, Piazza del Parlamento 1, 90134 Palermo,
	\and
	Centro de Astrobiolog\'{i}a / CSIC-INTA, LAEFF Campus, P.O. Box 78, 
     E-28691 Villanueva de la Ca\~nada, Madrid, Spain,  
      \and
	  ESO, Karl-Scharzschild-Strasse 2, D-85748 Garching-bei-M�nchen, Germany}
    \date{Received XXX; accepted XXX}


  \abstract
   {Different environmental conditions can play a crucial role in determining
final products of the star formation process and in this context, less favorable
activities of star formation are expected in the external regions of our Galaxy.}  
    {We   studied the properties of the young open cluster NGC\,1893 located about
12\,Kpc from the galactic center, to investigate how different physical conditions
can affect the process of star formation.}
    {By adopting a multiwavelength approach, we  compiled
a catalog extending from X-rays to NIR data to derive the cluster membership.
In addition, optical and NIR photometric
properties  are used to evaluate the cluster parameters.}
    {We find 415 diskless candidate members plus 1061 young stellar objects with a circumstellar disk or class\,II
candidate members, 125
of which are also H$_\alpha$ emitters.  Considering the
diskless candidate members, we find that  the cluster distance is   
3.6$\pm$0.2\,kpc and the mean interstellar reddening is E(B-V)=0.6$\pm$0.1 with evidence
of differential reddening in the whole surveyed region.}
    {NGC\,1893  contains a conspicuous population of pre-main sequence stars together 
with the well studied main sequence cluster
 population;  we  found a disk fraction of about 70\% similar to that 
found in clusters of similar age in the solar neighbour and then, despite expected  unfavorable conditions for star formation,
we conclude that  very rich young clusters can form also in the outer regions of our Galaxy. }
{}

\titlerunning{Star formation in the outer Galaxy: the young open cluster NGC1893}
    \maketitle
  %

  \section{Introduction}
The sequence of events that characterize the star formation (SF) process starts from the compression
of gas in a giant molecular cloud that produces molecular cores; these collapse into protostars
evolving in pre-main sequence (PMS) objects through accretion of material and circumstellar disk formation
\citep{zinn07}.
Within this standard picture, however, different environmental properties may play a key role  by changing 
the physical conditions during the SF process and possibly the final products, for example   the initial 
mass function (IMF). 

SF regions with peculiar properties are therefore crucial for a complete understanding 
of the SF process since they are the natural places where different initial conditions
can lead to the formation of clusters with non-standard IMF, dynamical evolution and/or
disk fraction and evolution.
In this context, clusters in the 
 outer Galaxy are very interesting since they are located in regions  where
surface and volume densities of atomic and molecular hydrogen are much smaller than in the inner Galaxy
\citep{wout90},
while metal content is, on average, smaller \citep{wils92}  
and therefore higher temperatures of the cloud are expected due to lower radiative losses.
In addition, the lack of prominent spiral arms and the presence of few supernovae as external triggers for SF
are further indications of less likely SF conditions. 

We have identified NGC\,1893 as a very promising cluster
for this kind of investigations since it is very young ($\sim$1-2\,Myr), distant about 12\,Kpc from the Galactic 
Center, with known massive members and indication of a  large PMS population \citep{vall99,shar07}.
This SF region, located at the center of the Aur OB2 associations, is associated with the HII region IC\,410 
and the two pennant nebulae, Sim\,19 and Sim\,130 \citep{gaze52}.  It is obscured by
several dust clouds and contains at least five O-type  stars \citep{hilt66}.  The presence of
several O-type  stars and 
emission line B-type stars \citep{marc02},
 likely in the PMS phase, is a strong indication
that very recent  SF events  have occurred in this cluster.

Several photometric and low resolution spectroscopic 
works have been devoted to this cluster since the first study by \citet{hoag61}.
Many of them are, however, aimed at studying the NGC\,1893 massive stars, that are in
the MS phase
\citep{beck71,moff72,cuff73,hump78,tapi91,fitz93,mass95,marc01,marc02}. Evidence
of a large population 
of PMS   stars in this cluster has been found by \citet{vall99} by using
near-infrared photometry,
even if their data did not allow them to distinguish the young population from
the contaminating
 field stars. Only recently,  \citet{shar07} presented a study about the low
mass population
in NGC\,1893 based on deep optical photometry down to V$\sim$22 and NIR 2MASS 
\citep{2mass} catalog. Several 
conclusions are drawn in this work about the IMF, the dynamical evolution and
the effects of massive cluster
stars on the low mass population. However, their results are based on a
statistical subtraction of the field
star population falling in the same region and suffer from the large statistical
uncertainties 
related to the membership.
In fact, the main challenge of the analysis of the low mass population in very
young 
clusters is the membership derivation since PMS stars usually lie
in the same region occupied by contaminating
field stars and it is therefore crucial an accurate analysis to identify cluster
members. 

We have started a project aimed at the  detailed study of the star formation  process in NGC\,1893
and in particular the unknown low mass population:
the first  results, based on 
the joint Chandra-Spitzer large program {\it The Initial
Mass Function in the Outer Galaxy: the star forming region NGC1893} (P.I. G. Micela), have been published in 
\citet[][ hereafter Paper I]{cara08}. In this latter paper,
 a conspicuous sample
of low mass  cluster members with circumstellar disk and of more evolved diskless candidate  members
 belonging to this region has been identified by using
Spitzer-IRAC and Chandra X-ray
data.

We present here the results on the membership and cluster parameters 
 based on a  multiwavelength approach including
 new deep optical and JHK photometry that  allow us
to derive a more complete census of the low mass population in this cluster;
knowledge of individual members is used to derive cluster parameters (reddening and distance)
in a more accurate way than previously reported in the literature. Detailed studies on the
X-ray coronal properties, disk fraction and evolution and IMF will be presented in forthcoming
papers.

The present paper is organized as follow. In Sect.\,2  we present the observations, the data reduction method and a description of the procedure used to derive
  the photometry and the astrometry. The whole catalog is presented in 
Sect.\,3 while    Sect.\,4 includes 
 the comparison with previous works;  Sect.\,5 describes  how 
we identify cluster candidate members by using VRIJHK, H$_\alpha$ and X-rays data, while in Sect.\,6
the cluster parameters are derived and a comparison with literature results is presented in Sect.\,7.
In Sect.\,8 individual masses and ages are derived while in Sect.\,9 we present our summary and conclusions.
  \section{Observations, data reduction and photometry}
We present here new optical and near infrared (NIR) data of the cluster,
 collected  from the Telescopio
Nazionale Galileo (TNG) at the Roque de los Muchachos Observatory (ORM, Canary
Islands, Spain), and from the 2.2\,m telescope at the Calar Alto Observatory (Spain).
  \subsection{Optical data}
 Optical observations were acquired in service mode using two different
telescopes, 
with the standard VRI  and H$\alpha$ photometric
filters: the Device Optimized for the LOw RESolution (DOLORES) mounted
on the TNG was used  in service mode during three nights in 2007, and 
the Calar Alto Faint Object Spectrograph  (CAFOS),  mounted on
the 2.2\,m telescope in Calar Alto German-Spanish Observatory (Spain),
observed for three nights in 2007 and 2008, as detailed in
Table~\ref{obslog}.  
Standard fields were taken with both instruments to perform the required
photometric calibration.
\begin{figure*}[!th]
   \centering
   \includegraphics[width=0.73\textwidth]{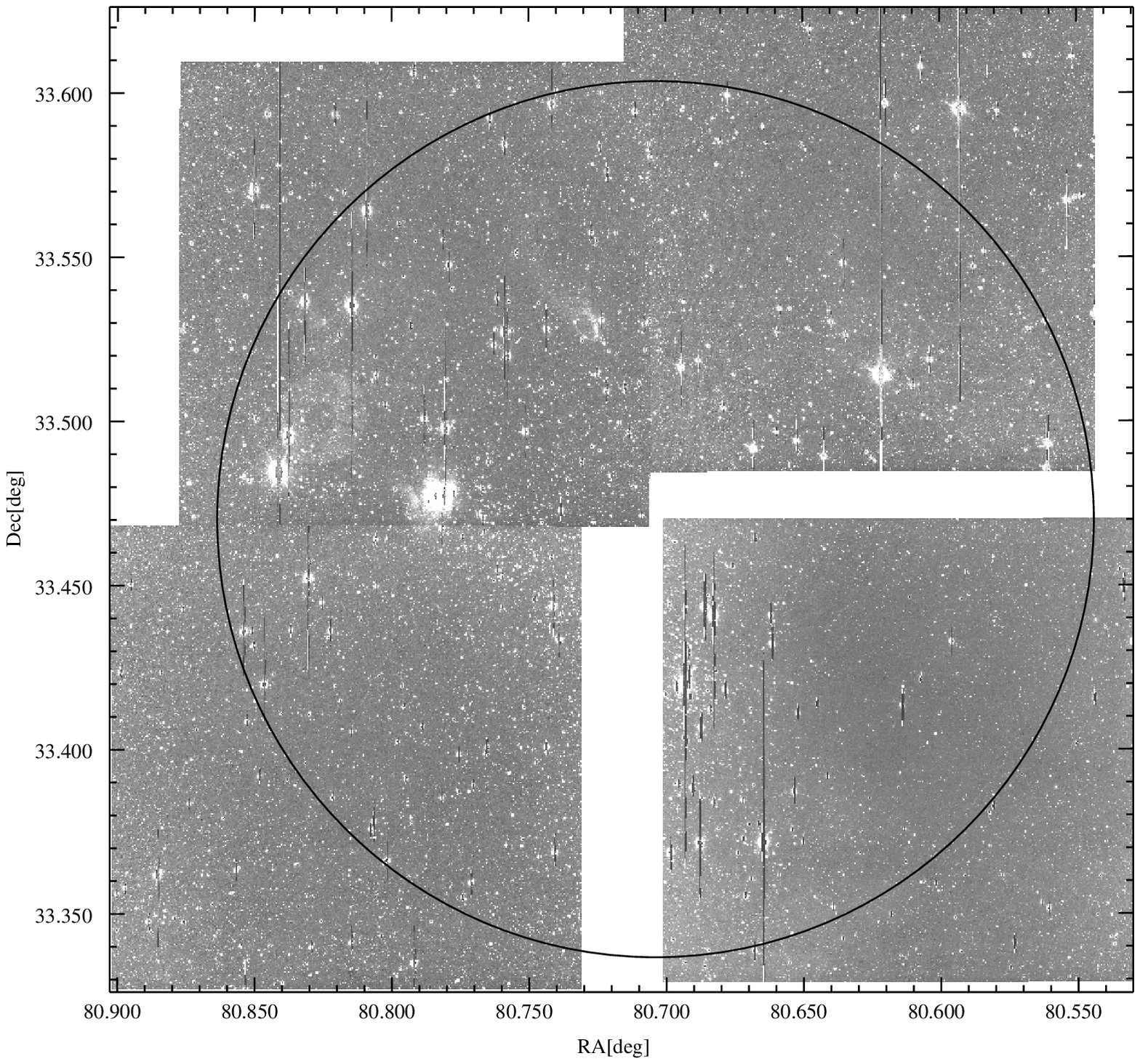} 
   \includegraphics[width=0.73\textwidth]{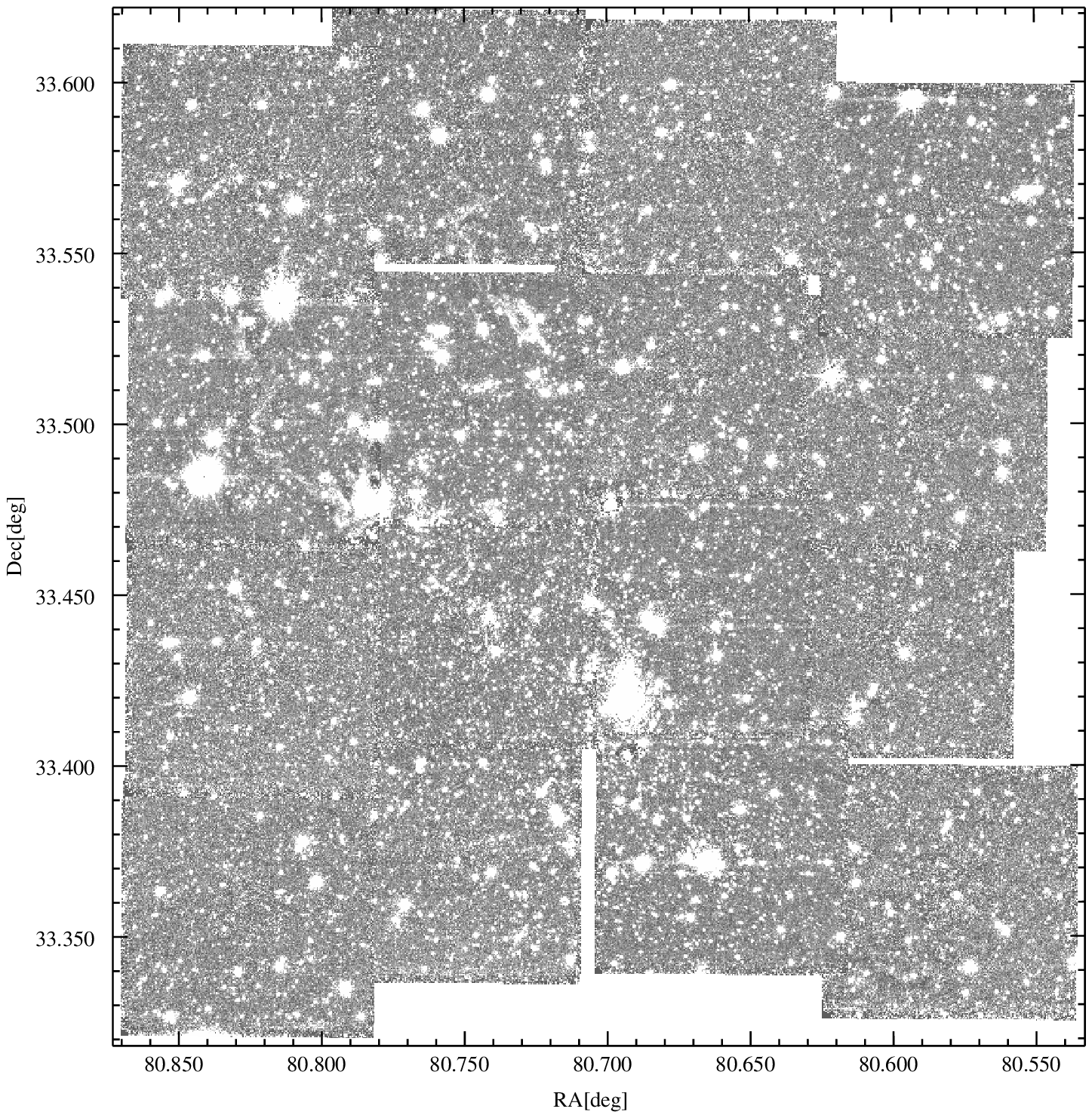}
   \caption{Upper panel:  I band combined image of the four DOLORES frames of 740\,sec; the overplotted
 circle indicates the CAFOS FoV.
Bottom panel: Js band combined image of the NICS frames  reduced and coadded with SNAP.}
   \label{images}
\end{figure*}

\begin{table*}
\tabcolsep 2.4pt
\caption{Optical observations of the cluster NGC 1893}\label{obslog}
\begin{center}
\begin{tabular}{lccccccc}
  \hline \hline
{Date}& $\alpha$ (J2000) & $\delta$ (J2000) & {Instrument}&{Filter}&{Exp. time[sec]} & {seeing} & {airmass} \\
  \hline
2007/09/21   & 5:22:27.4 & 33:23:46.0 & DOLORES  & V & 10, 60, $2\times 500$ & 1.0\arcsec & 1.151--1.192 \\
2007/09/21   & 5:22:27.5 & 33:23:36.9 & DOLORES  & R & 10, 70,          700  & 1.0--1.1\arcsec  & 1.066--1.079 \\
2007/09/21   & 5:22:27.4 & 33:23:42.5 & DOLORES  & I & 10, 60, $2\times 740$ & 0.9\arcsec & 1.095--1.138 \\
2007/09/21   & 5:22:27.6 & 33:23:33.0 & DOLORES  & H$\alpha$ & 60, 300, $2\times 700$ & 0.9--1.0\arcsec & 1.024--1.049 \\
2007/10/18   & 5:23:11.6 & 33:23:34.3 & DOLORES  & V & 10, 60, $2\times 500$ & 0.8\arcsec & 1.026--1.040 \\
2007/10/18   & 5:23:11.9 & 33:23:37.1 & DOLORES  & R & 10, 70,          700  & 0.8--1.0\arcsec & 1.006--1.010 \\
2007/10/18   & 5:23:11.8 & 33:23:22.0 & DOLORES  & I & 10, 60, $2\times 740$ & 0.8--1.0\arcsec & 1.009--1.021 \\
2007/10/18   & 5:23:12.8 & 33:23:47.2 & DOLORES  & H$\alpha$ & 60, 300, $2\times 700$ & 0.8\arcsec & 1.013--1.031 \\
2007/11/14   & 5:23:06.9 & 33:32:29.7 & DOLORES  & V & 10, 60, $2\times 500$ & 1.0\arcsec & 1.282--1.353 \\
2007/11/14   & 5:23:06.8 & 33:32:14.6 & DOLORES  & R & 10, 70,          700  & 1.0--1.1\arcsec & 1.113--1.134 \\
2007/11/14   & 5:23:06.7 & 33:32:24.5 & DOLORES  & I & 10, 60, $2\times 740$ & 0.8--0.9\arcsec & 1.184--1.257 \\
2007/11/14   & 5:23:06.7 & 33:32:08.2 & DOLORES  & H$\alpha$ & 20, 60, 300, $2\times 700$ & 0.8\arcsec & 1.050--1.096 \\
2007/11/14   & 5:22:28.7 & 33:32:30.5 & DOLORES  & V & 10, 60, $2\times 500$ & 1.0\arcsec & 1.004--1.007 \\
2007/11/14   & 5:22:33.9 & 33:32:39.3 & DOLORES  & R & $2\times 10$, 70, 700 & 0.8--0.9\arcsec & 1.015--1.022 \\
2007/11/14   & 5:22:32.0 & 33:32:20.5 & DOLORES  & I & 10, 60, $2\times 740$ & 0.9--1.0\arcsec & 1.004--1.010 \\
2007/11/14   & 5:22:34.2 & 33:32:47.4 & DOLORES  & H$\alpha$ & 60, $2\times 300$, $2\times 700$ & 0.8\arcsec & 1.029--1.066 \\
2007/10/11   & 5:23:22.0 & 33:26:38.7 & CAFOS & V & 15, $3\times 500$ & 1.4--1.6\arcsec & 1.002--1.004 \\
2007/10/11   & 5:23:21.5 & 33:26:37.4 & CAFOS & H$\alpha$ & 15, 100, $3\times 500$ & 1.6--1.8\arcsec & 1.002--1.016 \\
2008/01/05   & 5:23:23.3 & 33:26:42.4 & CAFOS & V & 10, 60, $2\times 300$ & 2.4\arcsec & 1.161--1.205 \\
2008/01/05   & 5:23:23.2 & 33:26:45.8 & CAFOS & R & 10, 60, $2\times 300$ & 2.3--2.6\arcsec & 1.231--1.283 \\
2008/01/05   & 5:23:23.3 & 33:26:47.3 & CAFOS & I & 10, 60, $3\times 500$ & 1.5--1.6\arcsec & 1.353--1.502 \\
2008/01/05   & 5:23:23.1 & 33:26:38.9 & CAFOS & H$\alpha$ & 10, 60, 500   & 1.4\arcsec & 1.603--1.651 \\
2008/01/09   & 5:23:24.2 & 33:26:41.6 & CAFOS & H$\alpha$ & 15, 100, $3\times 500$ & 1.2--1.7\arcsec & 1.139--1.239 \\
\hline
\end{tabular}
\end{center}
\end{table*}

DOLORES is a focal reducer instrument equipped with a 2048$\times$2048 
square pixels CCD, with a scale of 0.252\sec/px and a field of
view (FoV) of 8.6\arcmin$\times$8.6\arcmin. A mosaic of four fields
was made, covering an approximate area of
22.2\arcmin$\times$18.0\arcmin.
The standard VRI filters were used in the four fields, as well as
the H$\alpha$ narrow band filter centered on 6658\,$\AA$, with a Full Width
Half Maximum (FWHM) of 57\,$\AA$. We also observed the standard field SA\,98 
\citep[][$\alpha$=6$^{\rm h}$:52$^{\rm m}$:04.1$^{\rm s}$,
 $\delta$=-00$^{\rm d}$:19$^{\rm m}$:37.9$^{\rm s}$]{land92}.

CAFOS is a focal reducer instrument that works with a CCD  of
2048$\times$2048 square pixels, and has a scale of 0.53\sec/px,
covering a circular FoV of 16\arcmin\ diameter. 
With this FoV we covered the central areas of the cluster not  observed by 
 the four
DOLORES fields. Standard VRI filters were used, together with an H$\alpha$
filter centered at 6577\,$\AA$ with FWHM of 98\,$\AA$. The fields used as
standard for the photometric calibration were 
RU\,149 ($\alpha$=07$^{\rm h}$:24$^{\rm m}$:40.8$^{\rm s}$, 
	$\delta$=-00$^{\rm d}$:34$^{\rm m}$:10.0$^{\rm s}$), 
PG\,2331 ($\alpha$=23$^{\rm h}$:34$^{\rm m}$:14.1$^{\rm s}$,
	 $\delta$=+05$^{\rm d}$:46$^{\rm m}$:25.9$^{\rm s}$), 
and SA\,95 ($\alpha$= 03$^{\rm h}$:53$^{\rm m}$:32.9$^{\rm s}$, 
	    $\delta$=+00$^{\rm d}$:00$^{\rm m}$:48.3$^{\rm s}$). In this case we
noticed some variability in the photometric conditions of the last two
nights in Calar Alto. To improve the quality of the photometric
calibration we used also a set of local standards of the cluster,
selected during the TNG data calibration, as explained below.
A combined image of the four TNG fields in the
I band, with the shape of the CAFOS FoV over plotted, is shown in
Fig.\,\ref{images}. 

Optical data were reduced using the standard Image Reduction and
Analysis Facility (IRAF) routines, including the bias and flatfield
subtraction. Bias correction was applied by using Zero images and
the local overscan strip in each image. 
Some non-standard steps were taken due to various problems
 arisen during in the reduction, with details described below.

For the DOLORES observations
a combination of sky flat-fields was used to create a  master flat
field in each band, using the task \textsf{flatcombine}. Once the images were
corrected from flat-field, a strong illumination pattern was still
present in the images of the I band. An illumination function was
created using the flat-field (with the task \textsf{mkillumflat}), 
and conveniently removed from images. A
pattern of interference fringing was still present in the I band
images, with no variations found in the different nights. We
constructed a fringing image  by combining 15 images 
in the I band of the three nights, so the stars in the different
fields are canceled. The images selected were of
short exposure (2--60\,s) to reduce the number of stars in the
fields. The resulting fringing 
pattern was successfully corrected using the task \textsf{mkfringecor} and
applied to every image in the I band.

For the CAFOS data,
dome flats were used to correct the
images from local irregularities. The singular circular shape
of the FoV required special care: the values of all the
pixels outside  the FoV (easy to filter because they are well below the
background) were set to a very large number; in this way
no unreal features are created in the target images after subsequent
flat-field removal, resulting in a value well below background outside the
FoV. The  remaining  reduction followed the standard path.

\subsection{NIR data}
NIR observations were acquired in service mode at the TNG, using the
large field Near Infrared Camera Spectrometer (NICS), that is
a camera equipped with a  1024$\times$1024 square pixels HgCdTe HAWAII infrared array,
with a scale of 0.25\sec/pix and a FoV of 4.2\min$\times$4.2\min \citep{baff01}.
Our observations were acquired with   the
 Js(1.25\,$\mu$), H(1.63\,$\mu$) and K'(2.12\,$\mu$)
filters during eight nights in 2007 and 2008;
we observed the whole field with a raster of
   4$\times$4 pointings,  at each pointing we obtained a series of NINT dithered
       exposures; each exposure is a repetition of a DIT (Detector Integration Time) times
NDIT (number of DIT), to avoid saturation of the background;
details are given  in Table\,\ref{nirobslog}.

\begin{table*}
\tabcolsep 2.4pt
\caption{Near Infrared observations of the cluster NGC 1893 from NICS; NINTs is the number
 of single frames.}\label{nirobslog}
\begin{center}
\begin{tabular}{lcccccccc}
  \hline \hline
{Date}& {Field}&$\alpha$ (J2000) & $\delta$ (J2000) &{Filter}&{Exp. time[sec]} &{NINTs}& {seeing} & {airmass} \\
  \hline
2007/10/08  &  10  &  5:22:59.1    &  33:25:20.2    &       H    &   600   &   15    &     0.8\arcsec  &    1.103 \\
2007/10/08  &  10  &  5:22:59.0    &  33:25:23.3    &       J    &   510   &   17    &     0.8\arcsec  &    1.133 \\
2007/10/08  &  10  &  5:22:58.8    &  33:25:26.6    &       K'   &   700   &   14    &     0.8\arcsec  &    1.178 \\
2007/10/08  &  5   &  5:22:46.9    &  33:21:48.1    &       H    &   600   &   15    &     1.0\arcsec  &    1.003\\
2007/10/08  &  5   &  5:22:44.4    &  33:21:10.7    &       J    &   510   &   17    &     1.2\arcsec  &    1.004 \\
2007/10/08  &  5   &  5:22:41.5    &  33:21: 1.1    &       K'   &   700   &   14    &     0.8\arcsec  &    1.009 \\
2007/10/08  &  6   &  5:22:42.6    &  33:25:15.6    &       H    &   600   &   15    &     1.0\arcsec  &    1.022 \\
2007/10/08  &  6   &  5:22:42.1    &  33:25:20.0    &       J    &   510   &   17    &     1.0\arcsec  &    1.035 \\
2007/10/08  &  6   &  5:22:41.6    &  33:25:24.7    &       K'   &   700   &   14    &     1.0\arcsec  &    1.056 \\
2007/10/10  &  11  &  5:22:59.3    &  33:29:15.7    &       H    &   600   &   15    &     0.8\arcsec  &    1.197 \\
2007/10/10  &  11  &  5:22:59.1    &  33:29:18.6    &       J    &   510   &   17    &     1.0\arcsec  &    1.243 \\
2007/10/10  &  11  &  5:22:59.0    &  33:29:22.6    &       K'   &   700   &   14    &     0.8\arcsec  &    1.314 \\
2007/10/10  &  9   &  5:23: 0.2    &  33:21: 2.8    &       H    &   600   &   15    &     1.0\arcsec  &    1.073 \\
2007/10/10  &  9   &  5:22:60.0    &  33:21: 6.0    &       J    &   510   &   17    &     1.0\arcsec  &    1.096 \\
2007/10/10  &  9   &  5:22:59.8    &  33:21: 9.8    &       K'   &   700   &   14    &     1.0\arcsec  &    1.133 \\
2007/10/19  &  12  &  5:22:59.6    &  33:33:59.9    &       H    &   610   &   16    &     0.8\arcsec  &    1.467\\
2007/10/19  &  12  &  5:22:59.7    &  33:34: 0.0    &       J    &   510   &   17    &     0.8\arcsec  &    1.375 \\
2007/10/19  &  12  &  5:22:59.7    &  33:34: 0.0    &       K'   &   700   &   14    &     0.8\arcsec  &    1.313 \\
2007/10/19  &  15  &  5:23:17.2    &  33:28:55.8    &       H    &   600   &   15    &     0.9\arcsec  &    1.186 \\
2007/10/19  &  15  &  5:23:17.3    &  33:28:53.7    &       J    &   510   &   17    &     1.0\arcsec  &    1.144 \\
2007/10/19  &  15  &  5:23:17.5    &  33:28:50.5    &       K'   &   700   &   14    &     1.0\arcsec  &    1.114 \\
2007/10/19  &  16  &  5:23:19.0    &  33:32:58.1    &       H    &   600   &   15    &     1.0\arcsec  &    1.034 \\
2007/10/19  &  16  &  5:23:19.0    &  33:32:58.1    &       J    &   510   &   17    &     0.9\arcsec  &    1.049 \\
2007/10/19  &  16  &  5:23:18.7    &  33:34: 5.0    &       K'   &   710   &   15    &     0.9\arcsec  &    1.078 \\
2007/10/19  &  7   &  5:22:40.7    &  33:30: 5.1    &       H    &   600   &   15    &     0.9\arcsec  &    1.571 \\
2007/10/19  &  7   &  5:22:40.6    &  33:30: 5.1    &       J    &   510   &   17    &     0.9\arcsec  &    1.838 \\
2007/10/19  &  7   &  5:22:40.7    &  33:30: 4.8    &       K'   &   700   &   14    &     0.9\arcsec  &    1.707 \\
2007/12/17  &  1   &  5:22:21.9    &  33:21:15.0    &       H    &   600   &   15    &     1.2\arcsec  &    1.992 \\
2007/12/17  &  1   &  5:22:21.9    &  33:21:15.0    &       K'   &   700   &   14    &     1.2\arcsec  &    2.272 \\
2008/01/01  &  3   &  5:22:27.7    &  33:30:52.3    &       H    &   600   &   15    &     1.0\arcsec  &    2.047 \\
2008/01/01  &  3   &  5:22:27.6    &  33:30:52.5    &       J    &   510   &   17    &     1.0\arcsec  &    1.882 \\
2008/01/01  &  3   &  5:22:27.7    &  33:30:52.3    &       K'   &   700   &   14    &     1.2\arcsec  &    2.290 \\
2008/01/11  &  1   &  5:22:23.9    &  33:21:53.5    &       J    &   510   &   17    &     1.2\arcsec  &    1.419 \\
2008/01/11  &  4   &  5:22:23.9    &  33:33:53.5    &       J    &   510   &   17    &     1.5\arcsec  &    1.642 \\
2008/01/13  &  4   &  5:22:30.2    &  33:34: 5.0    &       H    &   650   &   16    &     1.0\arcsec  &    1.019 \\
2008/01/13  &  4   &  5:22:30.2    &  33:34: 4.9    &       K'   &   700   &   14    &     1.0\arcsec  &    1.034 \\
2008/01/16  &  13  &  5:23:25.9    &  33:22:22.1    &       H    &   600   &   15    &     0.8\arcsec  &    1.151 \\
2008/01/16  &  13  &  5:23:25.9    &  33:22:22.1    &       J    &   510   &   17    &     0.8\arcsec  &    1.196 \\
2008/01/16  &  13  &  5:23:25.9    &  33:22:22.1    &       K'   &   700   &   14    &     0.9\arcsec  &    1.248\\
2008/01/16  &  14  &  5:23:24.8    &  33:26:43.0    &       H    &   600   &   15    &     1.0\arcsec  &    1.586 \\
2008/01/16  &  14  &  5:23:24.8    &  33:26:42.9    &       J    &   510   &   17    &     0.8\arcsec  &    1.489 \\
2008/01/16  &  14  &  5:23:18.6    &  33:26: 5.8    &       K'   &   760   &   16    &     0.8\arcsec  &    1.342 \\
2008/01/16  &  2   &  5:22:20.4    &  33:25:10.2    &       H    &   138   &   31    &     1.2\arcsec  &    1.170 \\
2008/01/16  &  2   &  5:22:20.4    &  33:25:10.2    &       J    &   540   &   18    &     1.2\arcsec  &    1.213 \\
2008/01/16  &  2   &  5:22:18.4    &  33:25:38.5    &       K'   &   700   &   14    &     1.0\arcsec  &    1.047 \\
2008/01/16  &  8   &  5:22:39.8    &  33:32:47.8    &       H    &   560   &   14    &     1.2\arcsec  &    1.006 \\
2008/01/16  &  8   &  5:22:39.8    &  33:32:47.8    &       J    &   510   &   17    &     1.0\arcsec  &    1.011 \\
2008/01/16  &  8   &  5:22:39.8    &  33:32:47.8    &       K'   &   700   &   14    &     1.0\arcsec  &    1.023 \\
\hline
\end{tabular}
\end{center}
\end{table*}                           

Raw IR images taken with NICS were reduced by using the Speedy Near-IR data
 Automatic reduction Pipeline (SNAP) implemented  for NICS using the "full dither"
options. We forced the script to use an external master flat field obtained
by a median combination of single flat fields taken during the same night or the 
 nearest night.
After a bad pixel correction, the pipeline performs the source detection
on the sky subtracted and flat fielded images. A double iteration
is performed to obtain a coadded image for each filter
 by using the  offsets among the images computed by a
cross-correlation algorithm after the presence of the field
distortion is corrected.
A combined image in the Js band of the NICS images reduced and coadded with SNAP is shown in Fig.\,\ref{images}
 (bottom panel).


\subsection{Photometry and data selection \label{datasel}}
Sources detection and Point Spread Function (PSF) photometry were
 made using the routines within the Fortran stand-alone
versions of the DAOPHOT II and ALLSTAR packages \citep{ste87},
improved with the use of ALLFRAME \citep{ste94}. Instrumental
PSF photometry was corrected  by using appropriate  aperture 
corrections derived by growth curves computed with the DAOGROW code \citep{stet90}.
Details on the photometric calibration are given in Appendix.

  To perform an accurate analysis of the color-magnitude and color-color diagrams and compare them
  with appropriate theoretical models we selected the optical photometric catalogs obtained from the
 Dolores and Cafos images by considering only objects with photometric error smaller than a magnitude-dependent limit and with the DAOPHOT parameter 
{\tt sharp} smaller than 0.5 i.e. objects with brightness distribution consistent with point-like sources.
The  Dolores and Cafos final catalogs include 7144 and 3222 objects, respectively. 

  The photometric JHK catalog obtained using NICS images 
 was   selected by considering only objects with
  magnitude errors smaller than  
$\frac{0.2}{\sqrt2}$ in order to have color errors    smaller than 0.2.
 This criterion  was
adopted in order to discard false detections found in the spots due to saturated objects.
%
  After this selection  our JHK catalog  contained  
  11181 entries 
     including multiple items 
  for the objects  located in the overlapping regions of the 16 fields.  

\begin{figure}
    \centering
    \includegraphics[width=9cm]{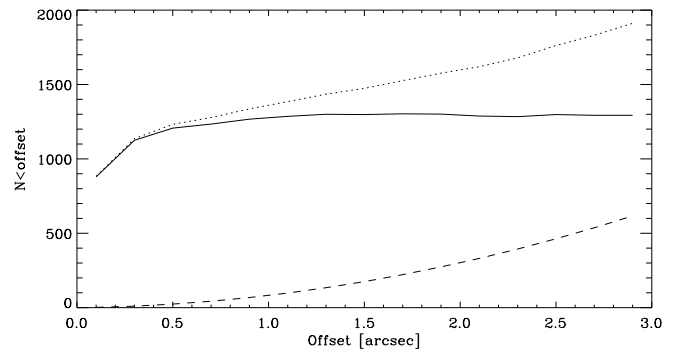}
      \caption{Cumulative distribution of offsets between objects falling in adjacent fields (dotted line).
	The expected offset cumulative distribution of spurious matches (dashed line, 
see Eq.\,\ref{spureq}) and the       {\it true} distribution (solid line, see text) are also shown.}
	  \label{nics_match_rad}
    \end{figure}

 To filter out the redundant
detections, we first  estimated an appropriate matching radius 
  by computing the offset cumulative distribution of the objects in adjacent fields 
within a relatively large 
radius equal  to 3\sec\ using a binsize of 0.2\sec,
as shown in Fig.\,\ref{nics_match_rad} (dotted line). 

We compared this distribution with 
  the expected cumulative distribution of spurious identifications (dashed line in Fig.\,\ref{nics_match_rad})
 computed  by using the relation 
  \begin{equation}
  \frac{dN}{dr}=\sum_{i=1,j=2}^{15,16}\frac{n_i*n_j}{A_{ij}}[\pi (r+dr)^2-\pi r^2]
\label{spureq}
  \end{equation}

  where $n_i$ and $n_j$ are the total number of objects in the $i$-th and $j$-th fields,
respectively, falling in the common area $A_{ij}$.
  The difference between the measured offset distribution and the expected spurious one 
  is the {\it true} distribution (solid line in Fig.\,\ref{nics_match_rad})
which tends to flatten at separations greater than 1.1\sec.

Using
  this value as  the final matching radius, we  included less than 10\% of spurious matches and  
  most of the true identifications. 

  Therefore, we matched each star of our catalog with those of adjacent
  fields and considered as the same object, those falling within 1.1\sec.
The final catalog composed of single items includes 
  9\,880 objects with K magnitude down to 17.78. 
%
%
\subsection{Astrometry}

In order to assign  celestial coordinates to the stars in the optical
catalogs, we used the 2MASS catalog \citep{2mass}
as reference. The transformation between the two coordinate systems
was made by applying the appropriate sky projection using several tasks of
the \textsf{imcoords} IRAF package. To determine this
transformation we first  identified through visual inspection 
three stars common to the 2MASS observations and to each of our
fields. With these initial coordinates we  used the task
\textsf{ccxymatch} to match the pixel coordinate list
of each of our fields and the celestial coordinate list 
of the 2MASS catalog. The   resulting single matched coordinate list
 is used as input of the task 
\textsf{ccmap} to derive the final sky projection. 
  This sky map is set in the master image of each field with the task
\textsf{ccsetwcs}, and the catalog of stars converted into celestial
coordinates using the \textsf{xy2sky} program of the \textsf{WCSTools}
UNIX   package  developed  at the Smithsonian Astrophysical Observatory. 
The assignment of celestial coordinates to the stars in the NIR catalog
 was performed by using the \textsf{WCSTools} package.

To verify the results and estimate the astrometric accuracy, we matched
 the three entire    catalogs DOLORES, CAFOS and NICS, with 
the 2MASS catalog, used as reference to find the astrometric
solution, and considered the offset distribution within a relatively
large value (2\sec). From this distribution we subtracted the expected
distribution of spurious identifications (Eq. 1),  and we obtained the distribution 
of {\it true} identifications.  For the matches DOLORES-2MASS, 
CAFOS-2MASS and NICS-2MASS, the 1\,$\sigma$ corresponding values
 are 0.24\sec, 0.22\sec\ and 0.19\sec, respectively,
 that are then
the final  accuracies for each catalog. 
  \section{Catalog}
  The aim of this section  is to describe the multiwavelength catalog obtained using the following data:
  \begin{itemize}
  \item{the new optical VRI photometry from DOLORES@TNG and CAFOS@2.2m telescope of the
Calar Alto Observatory presented here;}
  \item{the new deep JHK photometry from NICS@TNG telescope presented here;}
  \item{the Spitzer/IRAC photometry published in Paper I;} 
  \item{the Chandra/ACIS-I X-ray detections published in Paper I.}
  \end{itemize}

\begin{figure*}
    \centering
    \includegraphics[width=\textwidth]{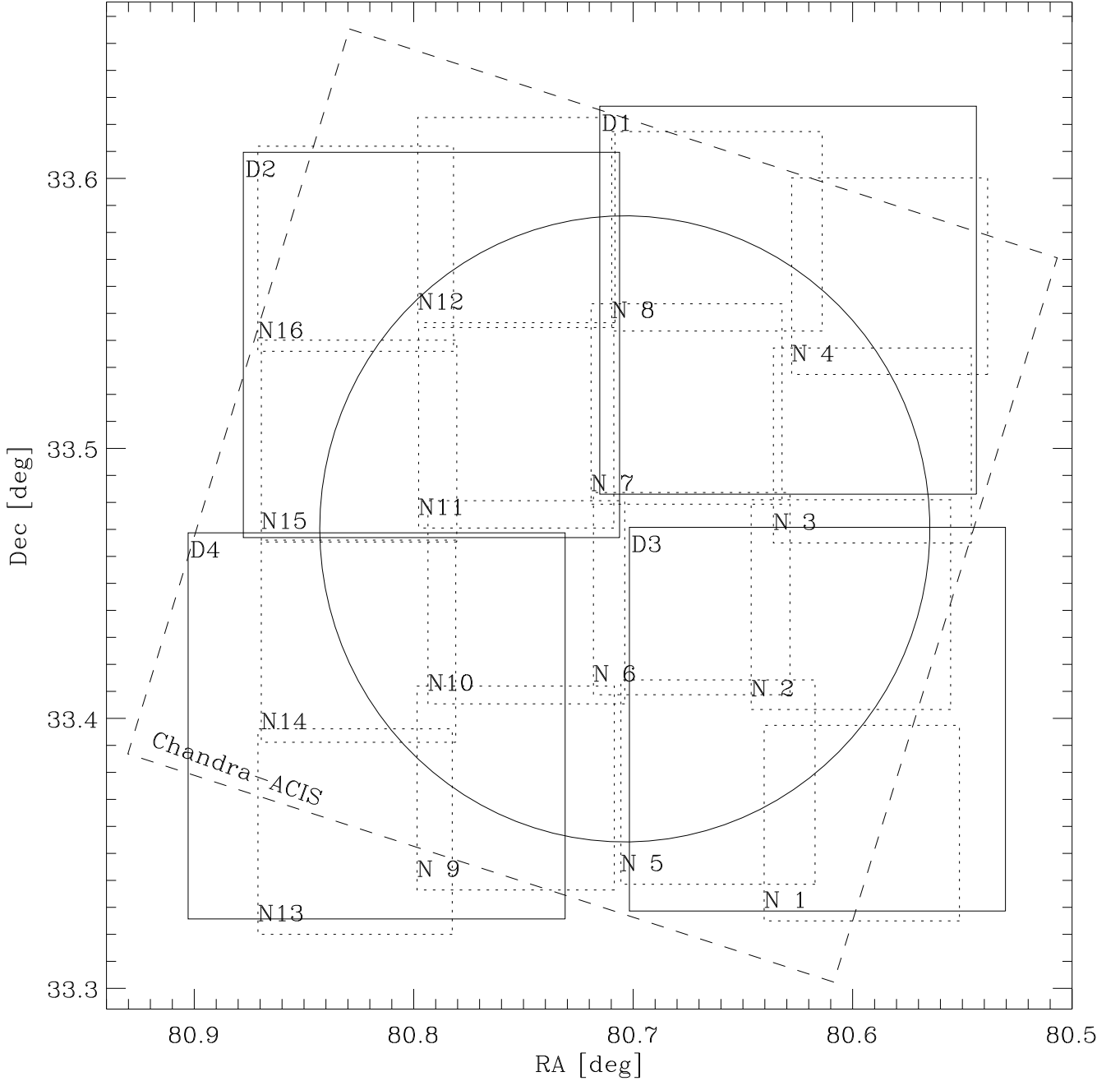}
      \caption{FoVs of our observations:  DOLORES (solid boxes), CAFOS (solid circle), 
NICS (dotted boxes) plus the Chandra-ACIS (dashed box). In the upper-left and bottom-left corners of the DOLORES (D) and NICS (N)
boxes, respectively, the corresponding field number is indicated.
 The IRAC FoV is larger  than all the previous ones and it is not  shown here. }
	  \label{observations_FoV_fig}
    \end{figure*}

  Figure\,\ref{observations_FoV_fig}
  shows the fields of view (FoV) of all involved observations, viz. DOLORES (4 solid boxes), 
CAFOS (solid circle), 
NICS (16 dotted boxes) plus the Chandra-ACIS\footnote{Note that the IRAC
  FoV is larger  than the fields shown in this figure.} (dashed box).
To match these catalogs we first
 derived an appropriate matching radius for each couple of catalogs, and then we
 identified the common objects by considering the nearest neighbor match within
the adopted radius.
 With the aim to include as many counterparts as possible  between the catalogs, we
initially computed 
  the offset distribution within 
  a radius larger than the expected astrometric accuracy of the two matched catalogs and the 
  distribution of the spurious identifications  according to the Eq. \ref{spureq}.
To match the objects we use the
radius   at which the {\it true} distribution (see Sect. \ref{datasel}) starts to 
flatten; in all cases we have
  less than 10\% of spurious identifications. 
  We derived a matching radius of 0.7\sec\,  between the NICS and IRAC catalogs.
  The IRAC catalog includes 15\,277 objects having at least one magnitude among the four 
IRAC filters; 7194 sources fall  within the NICS
  FoV,  5020 
    with a  counterpart 
  in the NICS catalog. 
  Most of the  NICS objects with no IRAC counterpart are fainter than  K=15,
  since NICS data are deeper than IRAC observations.
  Using the same method described before, we  found that a matching radius of 0.7\,\sec\ is appropriate
  to cross-match the DOLORES optical catalog with both the NICS and IRAC catalogs. 
  In the case of the DOLORES-NICS matches we  found 5285 
common objects;
  using the same radius, we cross-correlated the DOLORES objects without a NICS counterpart with the IRAC catalog
and we found further 629 objects, mostly located outside the NICS images,
for a total of 5914 DOLORES optical detections having at least one NICS or IRAC
counterpart.
  A radius of 0.7\,\sec\   was used also to cross-correlate the CAFOS  and 
NICS catalogs, finding 2915 common objects, 
 while a radius of 0.9\sec\ 
 was adopted to match the CAFOS unmatched objects  with the 
  IRAC catalog, finding further 64 objects.  

 The X-ray ACIS-I catalog  includes 1021 sources.  As mentioned in  Paper I, 717
of them have a single counterpart 
  in the IRAC catalog while 2 sources have a double identification.
  For the match
  between the NICS and X-ray ACIS-I catalogs we used the same procedure adopted
in Paper I, i.e. 
  we considered three different
  matching radii, 0.7\sec, 1.5\sec\, and 2.5\sec, in order to take into account
the degrading of the ACIS point spread function
  (PSF)  off axis.
  We  found a total of 752 X-ray sources  with a single counterpart in the NICS catalog 
and 10 X-ray sources with  two IR counterparts  
   642 having already an IRAC detection.

The entire catalog includes 21\,688 objects,  with 9531 having  only the IRAC counterpart
since the Spitzer/IRAC FoV is larger than that of the other observations.

 
%

\section{Photometry comparison}
We  compared the photometry obtained with DOLORES and CAFOS data
in order to check the consistency of our data.  We limited the comparison to V=22 since
the   limiting magnitudes in the CAFOS and DOLORES catalogs are V$\sim$22 and V$\sim24.3$, respectively,
if we consider magnitudes  with 
errors smaller than 0.1.
 The results of the statistical 
comparison between the catalogs are shown in
 Table\,\ref{photcomptab} where we also show the analogous results   obtained
 from the 
comparison with the catalog
of \citet{shar07}, that is the only VRI photometric catalog  available in literature.
For the latter comparison, we limited the \citet{shar07} catalog to V$<$20 in order to avoid effects
due to large photometric errors.
Our catalogs, obtained with CAFOS and DOLORES observations are in excellent agreement, as expected, since the
CAFOS one has been calibrated by using the DOLORES catalog. We  found a good agreement also with the \citet{shar07}
photometry since in all bands we have offsets smaller than 0.04 mag and the standard deviations
is always smaller than 0.2 mag. The mean offsets in the V-I color are all smaller than 0.05 mag.

A good agreement  is found also  from the comparison between our JHK photometry, obtained with NICS data, and
the 2MASS catalog, as  shown by the corresponding statistical values given in Table\,\ref{photcomptab}.

\begin{table}                                                          
\caption{Comparison between our and literature optical and NIR photometric catalogs.
 For each couple of   catalogs, statistical values and the number (N) of 
objects on which they have been computed are given.}\label{photcomptab}

\centering  
\tabcolsep 3.pt
\begin{tabular}{cccccc}
  \hline \hline
	      &		   &   Mean & $\sigma$& Median    & N \\
DOLORES-CAFOS & $\Delta V$ &   0.017 &   0.173  &  0.003   &         2336\\
"		 & $\Delta R$ &  -0.003 &   0.179  & -0.010   &         2336\\
"		 & $\Delta I$ &   0.013 &   0.209  &  0.003   &         2336\\
CAFOS-Sharma et al.   & $\Delta V$ &   0.024 &  0.135  &  0.024   &          409\\
"		  & $\Delta R$ &   0.044 &  0.153  &  0.034   &          412\\
"		   & $\Delta I$ &  -0.020 &  0.147  & -0.029   &          427\\
DOLORES-Sharma et al.& $\Delta V$ &   0.044 &  0.156  &  0.027   &          581\\
" & $\Delta R$ &   0.026 &  0.134  &  0.010   &          525\\
" & $\Delta I$ &  -0.010 &  0.172  & -0.037   &          464\\
NICS-2MASS	&$\Delta J$ &  -0.015 &    0.117 &   -0.034 &          707  \\
"		&$\Delta H$ &   0.026 &    0.118 &    0.006 &          737 \\
"	&$\Delta K$ &   0.005 &    0.107 &   -0.008 &          729  \\

 \hline
 \hline     
\end{tabular}
\end{table}

  \section{Member selection} 
In this section  we  
complement our optical photometric catalog by using the literature 
 V and I magnitudes by \citet{shar07} for the objects detected in X-rays with Chandra/ACIS-I  or 
in the NIR with NICS and/or IRAC   falling outside our optical fields or in saturated 
or corrupted regions of the V and I images. This  is only for the purpose to
 find an as complete as possible 
list of cluster candidate members, while in the next sections, where we derive
 cluster parameters, stellar masses and ages, we will use only our photometry 
 to be sure to have an homogeneous photometric system.

\subsection{Candidate members with circumstellar disk}
  Deep infrared observations in the JHK bands presented in this paper allow us to enlarge
  the sample of NGC\,1893 members published in Paper I, based on X-ray detection
  and/or infrared photometry in the four IRAC bands.

  As in  \citet{dami06} and in \citet{guar07,guar09} we
 selected YSOs with circumstellar disk by using reddening-independent indices
that allow us to distinguish objects reddened by the interstellar medium from 
those with IR excesses due to  the disk. 
We have defined a generic index as 
\begin{equation}
\label{qindex_eq}
Q_{\rm ABCD}={\rm (A-B)-\frac{E(A-B)}{E(C-D)}\times (C-D)}
\end{equation}
where (A-B) and (C-D) are two different colors while E(A-B)/E(C-D) depends only
on the adopted reddening law. In our case we used the reddening laws of \citet{muna96} and \citet{riek85},
for optical and  JHK bands, respectively,
since these are those that better follow our data in the color-color diagrams;
  for the IRAC bands we adopted the mean of the extinctions relative to the
 K band, derived for five SFRs in \citet{flah07}.
With the magnitudes  available in our
multiband catalog we have defined 
 20 reddening independent Q-indices  by using the couples of colors given in Table\,\ref{qindextab}.
These indices allow us to select in a way as complete as possible, all the objects
showing excesses  in a given color. For example, indices involving the (V-I) colors
should select YSOs with excess in the NIR band, if the (V-I) colors derive 
from the  photosphere.
However, we note that in case of accretion, the V band can be affected by the veiling
contribution; in this case, the Q index is the combination of excesses in two colors.
  
\begin{table}
\caption{Couples of colors used to define the reddening independent Q-indices as in
Eq.\,\ref{qindex_eq} and the limit
adopted to distinguish objects with normal photospheric colors from 
YSOs with excesses.}\label{qindextab}

\centering  
\begin{tabular}{ccccccc}
  \hline \hline
A-B & C-D & Q$_{\rm lim}$& & A-B & C-D & Q$_{\rm lim}$\\
 \hline
J - H  &  H - K 	&-0.017		& & J - K  &  K - [3.6]  &-1.350  \\
V - I  &  J - K 	&-0.472		& & J - K  &  K - [4.5]  &-1.012\\
V - I  &  J - H 	&-1.047		& & J - K  &  K - [5.8]  &-1.028\\
V - I  &  H - K 	&-0.044		& & J - K  &  K - [8.0]  &-1.462  \\
V - I  &  J - [3.6]	& -1.008	& & J - H  &  K - [3.6]  &-0.705  \\
V - I  &  J - [4.5]	&-0.908		& & J - H  &  K - [4.5]  &-0.493  \\
V - I  &  J - [5.8]	&-0.968 	& & J - H  &  K - [5.8]  &-0.503 \\
V - I  &  J - [8.0]	&-1.271		& & J - H  &  K - [8.0]  &-0.776  \\
\ldots	& \ldots	& 		& & H - K  &  K - [3.6]  &-0.644  \\
\ldots &  \ldots   	& 		& & H - K  &  K - [4.5]  &-0.519\\
\ldots &  \ldots    	&		& & H - K  &  K - [5.8]  &-0.525\\
\ldots &  \ldots    	&		& & H - K  &  K - [8.0]  &-0.686 \\
 \hline     
\end{tabular}
\end{table}
     
 For each reddening independent Q-indices we adopted a conservative limit
 that   chosen by considering the expected colors of main-sequence stars
   \citep{keny95} for the colors involving the VIJHK band. For the color 
involving the IRAC magnitudes [3.6], [4.5], [5.8] and [8.0] we defined the
limit by considering the Q-indices of the objects that in the    [3.6]-[4.5] vs. [5.8]-[8.0]
diagram have color compatible with zero within 2\,$\sigma$. 
By assuming these limits, given for each index in Table\,\ref{qindextab}, 
we consider objects with infrared excess  those  with 
  Q-index smaller within 3\,$\sigma$  than the adopted limit in a given index.

We note however that with the Q-index method all the selected objects 
have  a disk but  there is an ambiguity region  where stars  with disk cannot
be  distinguished by reddened objects and then the list of candidate  members with
 a disk can be incomplete.

%

With this criterion, we  selected 984\,class\,II YSOs; 
  by comparing the class\,II YSOs with the  
242 class\,II YSOs selected in Paper I 
we  found 187 
are common to the two samples, 
only 19 are class\,II according to the IRAC colors
but not according to our indices,  and 36 are class\,II according to the IRAC colors 
 that fall outside of the NICS FoV.

In our list of class\,II candidate members we  found 5 of the 7 YSOs classified in Paper I
as class\,0/I candidate members, for which we  maintained the  0/I classification and 
 discarded them 
 from our sample of class\,II candidate members.  
In summary, we have 1034 (984+19+36-5) class\,II YSOs of which
792 
are new candidate members with  a disk found in this work.  The latter include 17
 objects classified in Paper I as diskless candidate members: 
 we  checked that   the NICS photometry is of good quality 
and therefore we  classified these objects as class\,II candidate members.

\begin{figure}
    \centering
    \includegraphics[width=9cm]{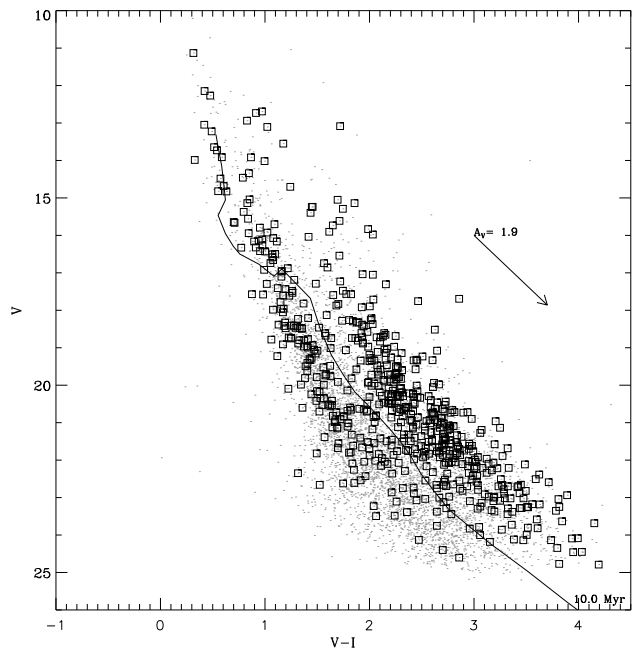}
     \caption{V vs. V-I  
diagram of all objects with optical magnitudes in our catalog.
 Class\,II YSOs are indicated with squares. 
Solid line is the 10\,Myr solar metallicity isochrone
of \citet{sies00} drawn assuming a distance of 3.6\,kpc  and E(B-V)=0.6.}
	  \label{viv_clII}
    \end{figure}

The V vs. V-I diagram of the objects with circumstellar disk 
is presented in Fig.\,\ref{viv_clII} where all class\,II candidate members are 
indicated by squares. The solid line is the 10\,Myr solar metallicity isochrone
of \citet{sies00},  drawn assuming a distance of 3.6\,kpc  and E(B-V)=0.6 (see Sect.\,\ref{paramsect});
 from now on the theoretical models will be plotted by using 
the \citet{muna96} and \citet{riek85}  reddening laws for the optical and NIR bands
and the \citet{keny95} conversions to transform theoretical temperatures and luminosities in the
observational plane.
The 10\,Myr isochrone is  used here only to distinguish the PMS region
from the  stars that have colors apparently inconsistent with a PMS nature. 

We  found that among the 1034 class\,II candidate  members, there are 170 YSOs in the region below the
 10\,Myr isochrone.  We do not believe these to be actually older than the other members, but rather, 
we guess they are candidate members for which optical colors are not purely photospheric. Most of these
objects show excesses in the JHK-IRAC colors and therefore the anomalous
optical colors and/or magnitudes could be due to phenomena related to the disk presence, 
such as veiling,
scattering or other effects due to the angle of inclination of the disk
 \citep[see ][for a discussion]{guar10}.

In fact, using the synthetic photometry derived from the \citet{robi06} models, 
 stars with circumstellar disk with age younger than 10\,Myr
 can have optical colors and magnitudes apparently compatible with  older age.
However we cannot  exclude that a small fraction of these objects could also be extragalactic sources,
as expected since the cluster NGC\,1893 lies in the direction of the  galactic anti-center.
We defer  to a  future
work an accurate analysis of the disk properties of these objects by means of
spectroscopic data.

%

  \subsection{Candidate  diskless members}
We define as diskless  candidate members all the objects 
having an X-ray counterpart in the Chandra-ACIS catalog,   showing any IR excess,
i.e.   not belonging to the  sample of class\,II YSOs, defined in the previous section,
and   falling in the PMS region of the V vs. V-I diagram, i.e. that with colors redder than
the 10\,Myr solar metallicity isochrone 
of \citet{sies00} drawn assuming a distance of 3.6\,kpc and E(B-V)=0.6 (see Sect.\,\ref{paramsect}).

\begin{figure}
    \centering
    \includegraphics[width=9cm]{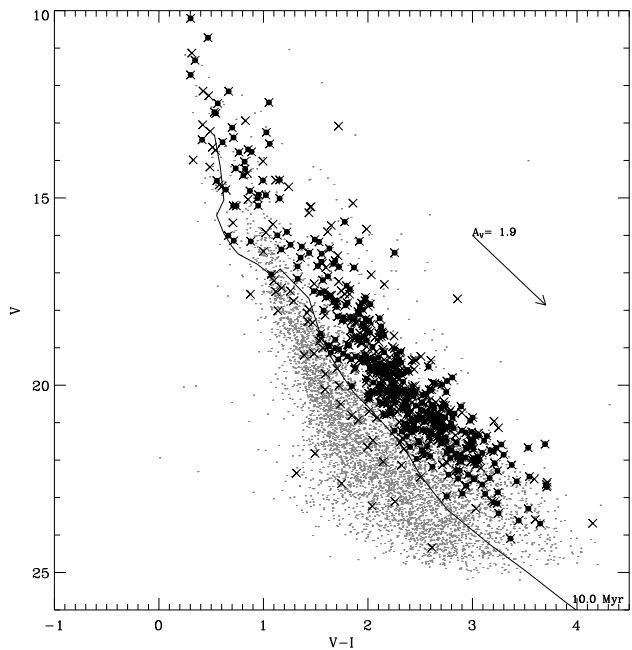}
      \caption{V vs. V-I diagram of all objects with optical magnitudes in our catalog.
X-ray detections are indicated with 
$\times$ symbols while those selected as diskless  YSOs are indicated with filled circles. Solid line is as in Fig.\,\ref{viv_clII}.}
	  \label{viv_clIII}
    \end{figure}

To define the PMS region, we have been guided by the bulk of objects
with X-ray counterpart, as shown in Fig.\,\ref{viv_clIII}  which depicts the V vs. V-I 
diagram of all objects with optical magnitudes in our catalog. Stars with X-ray counterpart
 are indicated with 
$\times$ symbols.  


In the sample of diskless candidate  members we also include the objects with X-ray emission that 
do not show any IR excess and have already reached the MS phase at the cluster age,
 i.e. those with V$<$15.

With the above conditions, we  found 434 
 diskless  YSOs, 85
 of which are in the sample of 110
diskless  YSOs found in Paper I. If we discard from the 110 diskless  YSOs defined in Paper I 
the 17 objects  considered here as class\,II YSOs,  
 the  remaining 8 diskless  YSOs classified
in Paper I  do not have either V or I magnitudes. By using the literature 
B and V magnitudes by \citet{mass95} of these 8 objects, we  found
 that 5  fall in the
cluster MS region while the other 3 lie in the PMS region: we therefore 
 included them in our sample 
of diskless cluster candidate members.

In summary, we have a total of 442 diskless  YSOs. 
These objects are indicated with filled circles in  the V vs. V-I diagram presented in Fig.\,\ref{viv_clIII}.

 \subsection{H$_\alpha$ emitters }
\begin{figure}
    \centering
    \includegraphics[width=9cm]{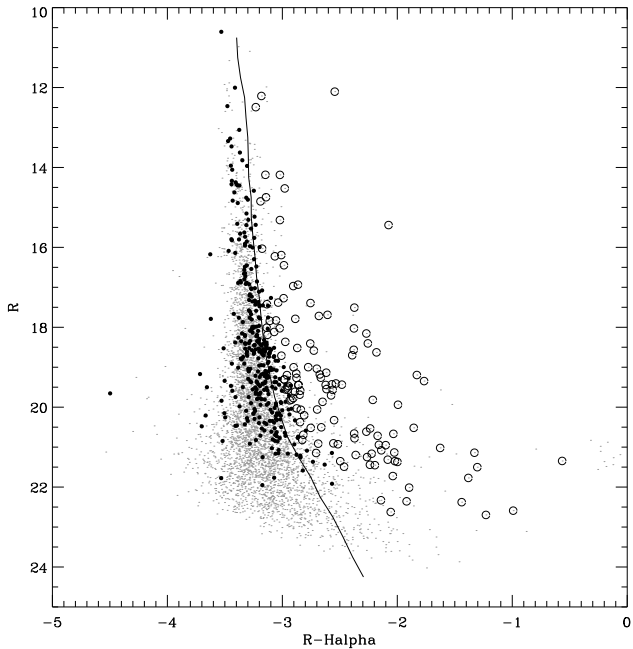}
      \caption{R vs. R-H$_\alpha$ diagram of all objects; 
 diskless  YSOs are indicated with filled circles while H$_\alpha$ emitters with at least
an other membership indication are indicated as empty circles. Solid line is the limit we defined
to distinguish diskless  YSOs  from high H$_\alpha$ emitters.}
	  \label{halpha_classification}
    \end{figure}

 In the case of stars
with accretion from circumstellar disk, the H$_\alpha$ line is much enlarged and therefore the 
(R-H$_\alpha$) index is a measure of the  H$_\alpha$ flux in 
excess. 
  In order to identify stars with H$_\alpha$ emission,
 we used the R vs. R-H$_\alpha$ diagram    shown
 in Fig.\,\ref{halpha_classification},
where  we defined the normal star limit by performing a linear fit of the R magnitudes  of all diskless  
YSOs (indicated with filled circles) having   R-H$_\alpha<-3$. Then we estimated a typical error in the
R-H$_\alpha$ index per bin of 0.5 magnitudes, as the average of the R-H$_\alpha$ errors in each bin.
We added such errors to the R-H$_\alpha$ derived from the linear fit to obtain the limit indicated in the
figure by the solid line.  We defined as H$_\alpha$ emitters, those with R-H$_\alpha$ larger within 3\,$\sigma$ than this limit. Among the 269 objects that satisfy this condition we 
consider as cluster H$_\alpha$ emitters only the 125 stars (indicated in the figure as empty circles)
 for which we have an additional membership 
 indication, viz. 98 class\,II and 27 objects previously classified as diskless candidate   members.

We  included in the class\,II sample these 27 H$_\alpha$ emitters,
 that we had classified as diskless  in the previous section.
  We note that this is not a contradiction in the classification but only a limit in the selection 
of  candidate  members with disk  since
there is a region in which the excesses by disks cannot
be distinguished from reddened objects.  

In summary, we have a sample of 1061 
 class\,II candidate  members, 125 of which are H$_\alpha$ emitters,
 plus 415\,diskless  stars.
Optical photometry and classification of class\,II and diskless candidates members are given in 
Tables\,\ref{optphotclII} and \ref{optphotclIII} while new JHK photometry and {\it Spitzer}-IRAC 
magnitudes from Paper I of class\,II and diskless candidates members are given in 
Tables\,\ref{nirphotclII} and \ref{nirphotclIII}\footnote{Full Tables\,\ref{optphotclII}, 
\ref{optphotclIII}, \ref{nirphotclII} and \ref{nirphotclIII} are also   
 available in electronic form at the CDS via anonymous ftp to
cdsarc.u-strasbg.fr  or via http://cdsweb.u-strasbg.fr}.
\tabcolsep 0.1truecm
\onltab{6}{
\begin{table*}
\caption{Optical photometry of Class\,0/I and Class\,II candidate members}
\label{optphotclII}
\centering
\begin{tabular}{cccccccc} 
\hline\hline
            Seq.   &            X ID   &        RA(2000)   &       Dec(2000)   &         V   &         R   &         I   &      H$_\alpha$  \\
            Num.   &                   &           [deg]   &           [deg]   &             &             &             &                  \\
\hline
     1    &      55    &      80.6220775    &      33.5139642    &   9.045$\pm$   0.016    &   9.241$\pm$   0.047    &                         &   0.036   \\
     2    &     779    &      80.7594680    &      33.5270160    &  11.132$\pm$   0.014    &  10.978$\pm$   0.017    &  10.817$\pm$   0.011    &   0.009   \\
     3    &     664    &      80.7418460    &      33.4434210    &  12.147$\pm$   0.014    &  11.923$\pm$   0.050    &  11.725$\pm$   0.016    &   0.009   \\
     4    &     276    &      80.6880800    &      33.4065610    &  12.269$\pm$   0.010    &  12.034$\pm$   0.035    &  11.793$\pm$   0.011    &   0.014   \\
     5    &     ...    &      80.7888715    &      33.5006385    &  12.692$\pm$   0.022    &  12.103$\pm$   0.007    &  11.721$\pm$   0.013    &   0.037   \\
     6    &     ...    &      80.6358343    &      33.5480553    &  12.733$\pm$   0.025    &                         &  11.819$\pm$   0.006    &           \\
     7    &     887    &      80.7798330    &      33.5475840    &  12.938$\pm$   0.011    &  12.491$\pm$   0.008    &  12.110$\pm$   0.014    &   0.009   \\
     8    &     345    &      80.6968720    &      33.4187280    &  13.047$\pm$   0.011    &                         &  12.628$\pm$   0.039    &           \\
     9    &     773    &      80.7583743    &      33.5198142    &  13.084$\pm$   0.015    &  12.211$\pm$   0.016    &  11.365$\pm$   0.021    &   0.007   \\
    10    &     ...    &      80.7442230    &      33.5280269    &  13.105$\pm$   0.007    &  12.588$\pm$   0.025    &  12.083$\pm$   0.022    &   0.006   \\
...\\
\hline
\end{tabular}
\begin{list}{}{}
\item[$^{\mathrm{}}$] Notes - This table is published in its entirety in the electronic edition of
Astronony \& Astrophysics. A portion is shown here for guidance regarding its form and content. 
\end{list}
\end{table*}
}

\tabcolsep 0.1truecm
\onltab{7}{
\begin{table*}
\caption{Optical photometry of diskless candidate members}
\label{optphotclIII}
\centering
\begin{tabular}{cccccccc} 
\hline\hline
            Seq.   &            X ID   &        RA(2000)   &       Dec(2000)   &         V   &         R   &         I   &      H$_\alpha$  \\
            Num.   &                   &           [deg]   &           [deg]   &             &             &             &                  \\
\hline
  1069    &     246    &      80.6833230    &      33.4407100    &  10.207$\pm$   0.042    &                         &   9.904$\pm$   0.011    &           \\
  1070    &     281    &      80.6885760    &      33.3713500    &  10.724$\pm$   0.009    &  10.605$\pm$   0.008    &  10.254$\pm$   0.020    &   0.035   \\
  1071    &     266    &      80.6865470    &      33.4434930    &  11.324$\pm$   0.034    &                         &  10.978$\pm$   0.013    &           \\
  1072    &      11    &      80.5768170    &      33.4725590    &  11.716$\pm$   0.019    &                         &  11.412$\pm$   0.038    &           \\
  1073    &       5    &      80.5541260    &      33.5671010    &  12.154$\pm$   0.012    &                         &  11.487$\pm$   0.025    &           \\
  1074    &     362    &      80.6991250    &      33.3685801    &  12.451$\pm$   0.006    &  12.004$\pm$   0.012    &  11.398$\pm$   0.014    &   0.041   \\
  1075    &     305    &      80.6923210    &      33.4224650    &  12.476$\pm$   0.020    &                         &  11.913$\pm$   0.022    &           \\
  1076    &     651    &      80.7396750    &      33.4333360    &  12.710$\pm$   0.016    &  12.465$\pm$   0.030    &  12.180$\pm$   0.019    &   0.006   \\
  1077    &     420    &      80.7068503    &      33.4259343    &  12.741$\pm$   0.018    &                         &  12.196$\pm$   0.011    &           \\
  1078    &     155    &      80.6625410    &      33.4408110    &  13.123$\pm$   0.022    &                         &  12.422$\pm$   0.006    &           \\
...\\
\hline
\end{tabular}
\begin{list}{}{}
\item[$^{\mathrm{}}$] Notes - This table is published in its entirety in the electronic edition of
Astronony \& Astrophysics. A portion is shown here for guidance regarding its form and content. 
\end{list}
\end{table*}
}

\tabcolsep 0.1truecm
\onltab{8}{
\begin{table*}
\caption{NIR photometry of Class\,0/I and Class\,II candidate members}
\label{nirphotclII}
\centering
\begin{tabular}{cccccccccc} 
\hline\hline
            Seq.   &        RA(2000)   &       Dec(2000)   &         J   &         H   &         K   &     [3.6]   &     [4.5]   &     [5.8]   &           [8.0]  \\
            Num.   &           [deg]   &           [deg]   &             &             &             &             &             &             &                  \\
\hline
     1    &      80.6220775    &      33.5139642    &                         &                         &                         &   8.290$\pm$   0.053    &   8.196$\pm$   0.036    &   8.179$\pm$   0.001    &   8.206$\pm$   0.036   \\
     2    &      80.7594680    &      33.5270160    &  10.468$\pm$   0.016    &                         &  10.583$\pm$   0.027    &  10.326$\pm$   0.041    &  10.219$\pm$   0.001    &  10.099$\pm$   0.004    &   9.718$\pm$   0.015   \\
     3    &      80.7418460    &      33.4434210    &  11.781$\pm$   0.043    &  12.814$\pm$   0.089    &  11.746$\pm$   0.019    &  11.045$\pm$   0.002    &  10.905$\pm$   0.002    &  10.802$\pm$   0.005    &  10.662$\pm$   0.016   \\
     4    &      80.6880800    &      33.4065610    &  11.574$\pm$   0.016    &  11.519$\pm$   0.019    &  11.306$\pm$   0.008    &  11.331$\pm$   0.002    &  11.255$\pm$   0.002    &  11.373$\pm$   0.017    &  11.399$\pm$   0.051   \\
     5    &      80.7888715    &      33.5006385    &  10.905$\pm$   0.039    &  10.434$\pm$   0.029    &  10.116$\pm$   0.018    &   9.262$\pm$   0.041    &   8.894$\pm$   0.042    &   8.568$\pm$   0.002    &   8.380$\pm$   0.003   \\
     6    &      80.6358343    &      33.5480553    &  11.859$\pm$   0.070    &  11.430$\pm$   0.059    &  11.477$\pm$   0.061    &  10.686$\pm$   0.001    &  10.747$\pm$   0.001    &  10.726$\pm$   0.005    &  10.811$\pm$   0.025   \\
     7    &      80.7798330    &      33.5475840    &  11.403$\pm$   0.036    &  11.630$\pm$   0.048    &  11.099$\pm$   0.036    &  10.927$\pm$   0.001    &  10.959$\pm$   0.001    &  10.963$\pm$   0.009    &  10.992$\pm$   0.026   \\
     8    &      80.6968720    &      33.4187280    &  12.439$\pm$   0.016    &  12.501$\pm$   0.034    &  12.326$\pm$   0.013    &  11.992$\pm$   0.016    &                         &  11.971$\pm$   0.036    &  11.845$\pm$   0.079   \\
     9    &      80.7583743    &      33.5198142    &                         &                         &                         &   9.367$\pm$   0.041    &   9.376$\pm$   0.048    &   9.295$\pm$   0.002    &   9.348$\pm$   0.010   \\
    10    &      80.7442230    &      33.5280269    &                         &                         &                         &  10.697$\pm$   0.002    &  10.654$\pm$   0.002    &  10.517$\pm$   0.017    &  10.015$\pm$   0.054   \\
...\\
\hline
\end{tabular}
\begin{list}{}{}
\item[$^{\mathrm{}}$] Notes - This table is published in its entirety in the electronic edition of
Astronony \& Astrophysics. A portion is shown here for guidance regarding its form and content. 
\end{list}
\end{table*}
}

\tabcolsep 0.1truecm
\onltab{9}{
\begin{table*}
\caption{NIR photometry of diskless  candidate members}
\label{nirphotclIII}
\centering
\begin{tabular}{cccccccccc} 
\hline\hline
            Seq.   &        RA(2000)   &       Dec(2000)   &         J   &         H   &         K   &     [3.6]   &     [4.5]   &     [5.8]   &           [8.0]  \\
            Num.   &           [deg]   &           [deg]   &             &             &             &             &             &             &                  \\
\hline
  1069    &      80.6833230    &      33.4407100    &                         &   9.394$\pm$   0.053    &   9.532$\pm$   0.043    &   9.559$\pm$   0.027    &   9.605$\pm$   0.040    &   9.580$\pm$   0.003    &   9.658$\pm$   0.010   \\
  1070    &      80.6885760    &      33.3713500    &  10.041$\pm$   0.021    &   9.906$\pm$   0.019    &                         &   9.840$\pm$   0.039    &   9.914$\pm$   0.001    &   9.903$\pm$   0.002    &   9.924$\pm$   0.009   \\
  1071    &      80.6865470    &      33.4434930    &  10.545$\pm$   0.025    &  10.508$\pm$   0.018    &  10.628$\pm$   0.012    &  10.582$\pm$   0.002    &  10.574$\pm$   0.002    &  10.637$\pm$   0.009    &  10.697$\pm$   0.029   \\
  1072    &      80.5768170    &      33.4725590    &  11.348$\pm$   0.035    &  11.103$\pm$   0.035    &  10.982$\pm$   0.014    &  10.957$\pm$   0.001    &  10.946$\pm$   0.001    &  10.954$\pm$   0.006    &  11.029$\pm$   0.023   \\
  1073    &      80.5541260    &      33.5671010    &  10.811$\pm$   0.020    &  10.577$\pm$   0.027    &  10.545$\pm$   0.010    &  10.506$\pm$   0.035    &  10.449$\pm$   0.001    &  10.503$\pm$   0.005    &  10.539$\pm$   0.017   \\
  1074    &      80.6991250    &      33.3685801    &                         &                         &                         &  10.259$\pm$   0.070    &  10.395$\pm$   0.001    &  10.375$\pm$   0.004    &  10.370$\pm$   0.014   \\
  1075    &      80.6923210    &      33.4224650    &  11.485$\pm$   0.017    &  11.361$\pm$   0.019    &  11.266$\pm$   0.009    &  11.316$\pm$   0.014    &  11.289$\pm$   0.071    &  11.259$\pm$   0.030    &  11.136$\pm$   0.062   \\
  1076    &      80.7396750    &      33.4333360    &  11.667$\pm$   0.022    &  11.442$\pm$   0.023    &  11.349$\pm$   0.012    &  11.177$\pm$   0.002    &  11.143$\pm$   0.002    &  11.160$\pm$   0.007    &  11.102$\pm$   0.022   \\
  1077    &      80.7068503    &      33.4259343    &  12.128$\pm$   0.094    &  12.254$\pm$   0.150    &  11.851$\pm$   0.105    &  11.481$\pm$   0.004    &  11.512$\pm$   0.005    &  11.302$\pm$   0.049    &  11.431$\pm$   0.040   \\
  1078    &      80.6625410    &      33.4408110    &  11.902$\pm$   0.017    &  11.711$\pm$   0.018    &  11.628$\pm$   0.010    &  11.494$\pm$   0.001    &  11.490$\pm$   0.002    &  11.474$\pm$   0.009    &  11.497$\pm$   0.023   \\
...\\
\hline
\end{tabular}
\begin{list}{}{}
\item[$^{\mathrm{}}$] Notes - This table is published in its entirety in the electronic edition of
Astronony \& Astrophysics. A portion is shown here for guidance regarding its form and content. 
\end{list}
\end{table*}
}

  \section{Cluster parameters\label{paramsect}}

  Metallicity, interstellar reddening and cluster distance are fundamental to 
  compare the observed CMD with theoretical 
  tracks and isochrones from which we estimate masses and ages of cluster candidate members.
 However, as already discussed in several papers \citep[e.g.][]{hill97,gull98}, class II YSOs typically 
  experience  accretion phenomena, scattering and/or reflection processes
  due to the dust grains in their circumstellar disks that may alter
  their photospheric colors, also in the optical bands. In such cases
  comparison of observed magnitudes and colors with  models can be unreliable
  and the results should be interpreted   with caution. In order to get around
  this problem, we derive cluster parameters by considering only cluster candidate members without
  circumstellar disk (class III YSOs) for which the observed colors
  can be directly compared with evolutionary tracks and isochrones once the interstellar 
  reddening is taken into account. 

  \subsection{Metallicity}
In this work we adopt models computed for solar metallicity, since there are
ambiguous indications in the literature about the metallicity in NGC\,1893. In a paper focused on  
 the Galactic metallicity, \citet{roll93} and \citet{roll00} 
derived C, N, O, Mg, Al and Si abundances of 80 B-type main sequence stars, 8 of
which are NGC\,1893
members. After discarding two high rotator objects,  they performed an LTE
analysis  from which
 they derive for NGC\,1893 only a slight indication of under solar abundance. 
More recently, \citet{dafl04}
derived non-LTE  chemical abundances in young OB stars from high-resolution
echelle spectra.
 The two NGC\,1893 cluster members show again a marginal indication of subsolar
metallicity, but
from the chemical analysis of 69 young stars with Galactocentric distances
between 4.7 and 13.2\,kpc, 
\citet{dafl04} find  metallicity slopes consistent with a flattening of the
radial gradient. Even if  \citet{roll00} 
found indication of  steeper slopes, the   available studies do not show strong
evidence of sub-solar metallicity and therefore, as assumed in previous studies,
we adopt  a solar metallicity for NGC\,1893.

  \subsection{Interstellar reddening}
\begin{figure}
    \centering
    \includegraphics[width=9cm]{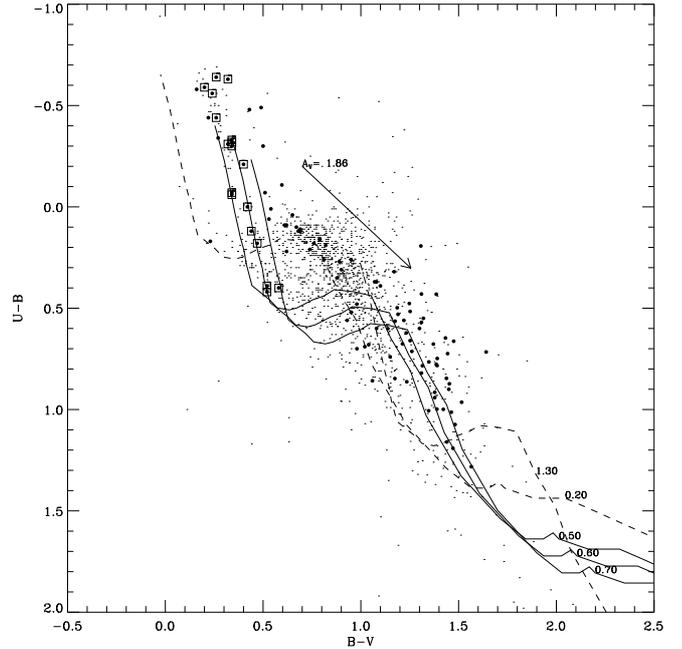}
      \caption{U-B vs. B-V diagram obtained by using \citet{mass95} and \citet{shar07} photometry  (dots);
  filled circles are diskless  candidate members, squares are those compatible
with reddening $0.5<E(B-V)<0.7$  based on the
  expected photospheric colors  by \citet{keny95} drawn  at the   E(B-V) values indicated on each line.
  The reddening vector obtained using the \citet{muna96} reddening law is also shown.}
	  \label{reddeningfig3}
    \end{figure}

  Color-color diagram using UBV photometry is the classical tool to derive 
  the mean interstellar reddening of a given cluster by comparing expected photospheric
  colors with those observed for diskless cluster members. 
  Fig.\,\ref{reddeningfig3} shows the U-B vs. B-V diagram obtained by using literature UBV photometry
  by \citet{mass95} and \citet{shar07}; diskless  stars  are indicated with filled circles. We find that most of the bluest diskless  YSOs,
indicated by empty squares, are compatible with
  a reddening E(B-V)=0.6$\pm$0.1 from the comparison with the
   solar metallicity ZAMS of \citet{sies00}, transformed into the observed colors
 by using the \citet{keny95} relations.
  Diskless  candidate members with B-V$\gtrsim$0.6 can be compatible with lower reddenings, down to
  E(B-V)=0.2, or with higher reddenings, at least up to E(B-V)=1.3, as shown by the ZAMS reddened with E(B-V)=0.2 
  and E(B-V)=1.3 (dashed lines in Fig.\,\ref{reddeningfig3}).

  This allows us to deduce that the cluster region is  affected by a significant differential reddening and 
  that for the reddest objects the interstellar reddening cannot be photometrically derived. 
On the contrary, it 
  can be unequivocally derived for  the bluest objects, i.e. those indicated with squares for which 
the $0.5<$E(B-V)$<0.7$. This result has been checked by computing individual reddenings for 
  the 12 diskless YSOs with U-B$\lesssim0.5$ 
  for which spectral types are available  \citep{hilt66,mass95,marc02}; we 
 found that among the bluest
  objects (B-V$\lesssim$0.5), all the 
  reddening values are compatible with  E(B-V)=0.6$\pm$0.1, while among the reddest ones 
(B-V$\gtrsim$0.5),  four have  E(B-V)$\sim$0.2
  and one has E(B-V)=0.68.

\begin{figure}
    \centering
    \includegraphics[width=9cm]{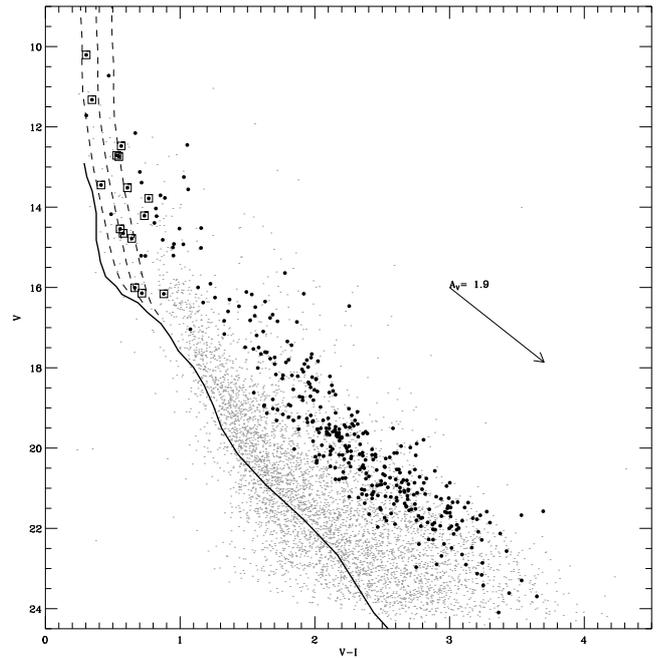}
      \caption{V vs. V-I diagram of all the objects (dots) with errors in V-I smaller than 0.1.
  Filled circles are the diskless candidate  cluster members while squares indicate those with E(B-V) between 0.5 and 0.7,
  according to Fig.\,\ref{reddeningfig3}.
Solid line is the solar metallicity ZAMS of \citet{sies00} at a distance of 3.6\,kpc and reddening E(B-V)=0.4. 
    Dashed lines indicate the  1.5\,Myr \citet{mari08} isochrone of solar metallicity for masses larger than 2\,M$_\odot$ at 3.6\,kpc and for 
  E(B-V)=0.5, 0.6 and 0.7. The reddening vector obtained using the \citet{muna96} reddening laws is also shown.}
	  \label{reddeningfig2}
    \end{figure}

  The mean reddening E(B-V)=0.6$\pm$0.1 we derive for NGC\,1893  is consistent also in 
  the V vs. V-I CMD shown in Fig.\,\ref{reddeningfig2}.
  In this figure, dots indicate all objects in the FoV with error in V-I smaller than 0.1 and filled 
  circles indicate diskless candidate  cluster members. Objects indicated also with squares are those with E(B-V)
  between 0.5 and 0.7, as in Fig.\,\ref{reddeningfig3}. 
  The thick line is
  the solar metallicity    ZAMS  of \citet{sies00}
  at a distance of 3.6\,kpc (see Sect.\,\ref{distancesect}) and
  reddened using  E(B-V)=0.4.   The value E(B-V)=0.4 is that assumed for foreground
  objects that we expect to be less reddened than cluster members.
To fit the highest mass range not covered by the \citet{sies00} ZAMS, we use
  the  theoretical isochrone at solar metallicity of 1.5\,Myr  of \citet{mari08}
  reddened with E(B-V)=0.5, 0.6 and 0.7 at a distance of  3.6\,kpc (dashed lines).
  In this diagram, diskless  candidate cluster members show  a large spread that can be due to several reasons, 
such as  differential reddening, binarity and/or age spread.   This is true also in the 
  V$\lesssim16$ range where we can have objects in PMS or MS phase. However, 
those with 0.5$<$E(B-V)$<$0.7, as derived from Fig.\,\ref{reddeningfig3} (indicated 
  with squares)  
  are well fitted by the  1.5\,Myr isochrone if a reddening E(B-V)=0.6$\pm$0.1 is adopted, in agreement
  with the value we derived 
  by using the U-B vs. B-V diagram.
  On the contrary, the other bright objects show a significant  spread and 
  can be both PMS objects or very reddened MS members.
  We note also that the  choice of the 1.5\,Myr isochrone is not critical for this result since
  in this mass range,  isochrones degenerate  in a vertical shape branch. For this reason in the V vs. V-I
  diagram, MS members cannot be used to constrain neither ages, nor  the cluster distance, while 
they can constrain the cluster reddening.

\begin{figure}
    \centering
    \includegraphics[width=9cm]{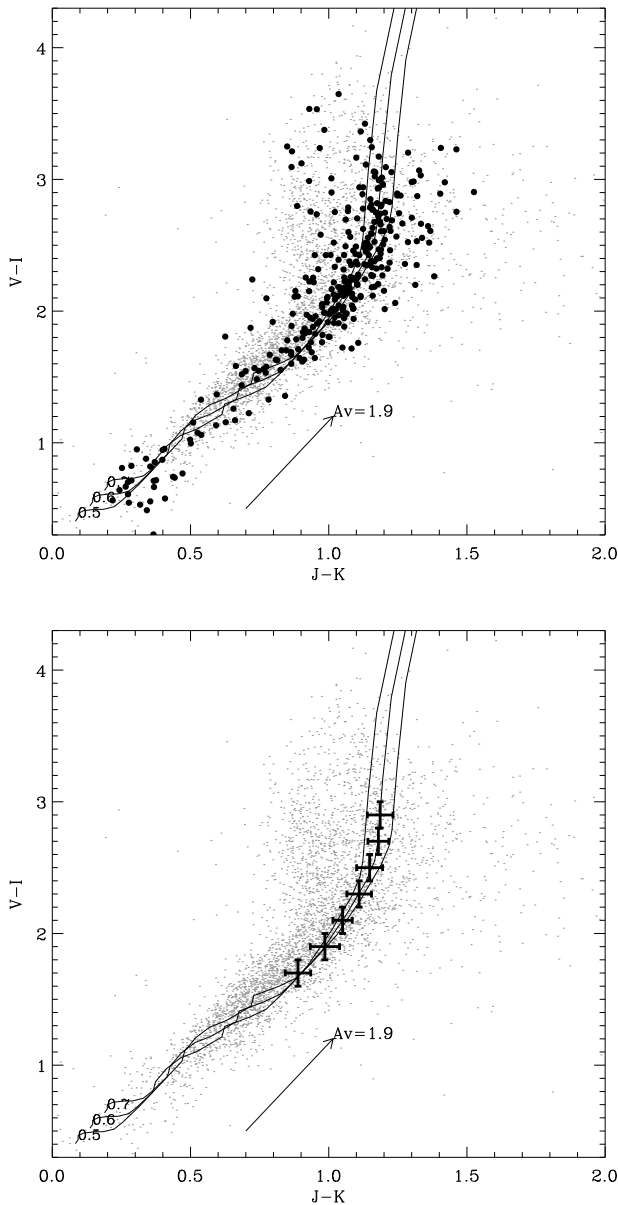}
      \caption{Color-color diagram of objects with error in color smaller than 0.1 (dots);
  solid lines are the photospheric colors  by \citet{keny95}   at the three  E(B-V) values indicated on each line.
  Filled circles in the upper panel are the NGC\,1893 candidate cluster
  members without circumstellar disk while error bars in the bottom panel indicate the cluster
  sequence fiducial line derive as described on the text. The reddening vector obtained using the \citet{muna96} and \citet{riek85} reddening laws is also shown.}
	  \label{reddeningfig1}
    \end{figure}

  Our result is also consistent in the low mass range, as can be seen by using
the V-I vs J-K diagram shown in 
  Fig.\,\ref{reddeningfig1}, where we  present
    the color-color diagram of all objects;
diskless  candidate members are used to estimate
the cluster reddening by considering
 the locus of expected photospheric colors by
\citet{keny95}; we take advantage  of the fact  that in this diagram  the 
expected photometric colors for
low mass stars are almost vertical.

  In order to estimate the cluster reddening, we considered also the
distribution
  of the J-K colors of diskless candidate  members within fixed ranges of 0.2 magnitudes
in the interval $1.3<$V-I$<3.0$,
 and we considered, for each V-I bin, the peaks of the J-K distribution to
derive the fiducial cluster sequence, 
indicated by the error bars  in the bottom panel of Fig.\,\ref{reddeningfig1}.
Comparison of this fiducial 
sequence with the lines allows us to confirm that the NGC\,1893 mean reddening
is E(B-V)=0.6$\pm$0.1.  However, since
the population of low mass stars is larger and spatially spread over the whole
region  affected by a larger 
differential reddening, we note that low mass candidate cluster members can individually
have E(B-V)  larger or smaller  values.  As it will be shown later,  by using
 IR colors we estimate that
the   reddening in this region is not larger than about E(B-V)=1.3 (A$_{\rm V}$=4 mag). 

The spread of low mass candidate cluster members in the V vs. V-I diagram is not due only to differential reddening. In fact,
if we select the sample of low mass diskless candidate 
 members from the V-I vs J-K diagram with 0.5$<$E(B-V)$<0.7$, we find that 
in the V vs. V-I diagram they show the same spread as all diskless  and therefore we conclude that 
  the low mass star spread in the V vs. V-I diagram could  
be also due  to an intrinsic age spread and/or binarity.

   We note that the E(B-V) value we derived both for low and high mass candidate  cluster
members, is independent
   of other adopted parameters such as distance, ages and metallicity.
However further spectroscopic  observations should be obtained to derive
spectral types and therefore individual
 reddening values.
 
  \subsection{Distance\label{distancesect}}
  The distance of NGC\,1893 is a very uncertain parameter ranging in the literature 
from 3.2 to 6.0\,kpc (see Sect. \ref{compareparam}).
 Using broad band optical or NIR photometry of candidate  cluster members and the isochrone fitting method,
  it is very difficult to derive the cluster distance. In fact,
  very young clusters such as NGC\,1893 include a population of PMS stars, that lies in a ample
  region of the CMD diagram, and a small population, more or less  rich, depending on the cluster age, of higher mass objects already on the MS phase and therefore on a well defined locus of the CMD.
  Nevertheless, the MS star distribution on the CMD is usually almost vertical and therefore quite ineffective
 for an accurate  distance derivation. In addition,
  for clusters younger than about 15\,Myr, the flattening of the isochrones between the two phases (PMS and MS)
  ends on the MS at a given magnitude that is strongly dependent on the isochrone age. Unfortunately,
  unless individual ages of PMS cluster members   are well known,
  it is  very hard to constrain the cluster distance by looking for this magnitude,
  even if the cluster mean reddening is fixed and a large population of high and low mass cluster members
  is known.

\begin{figure}[!t]
    \centering
    \includegraphics[width=9cm]{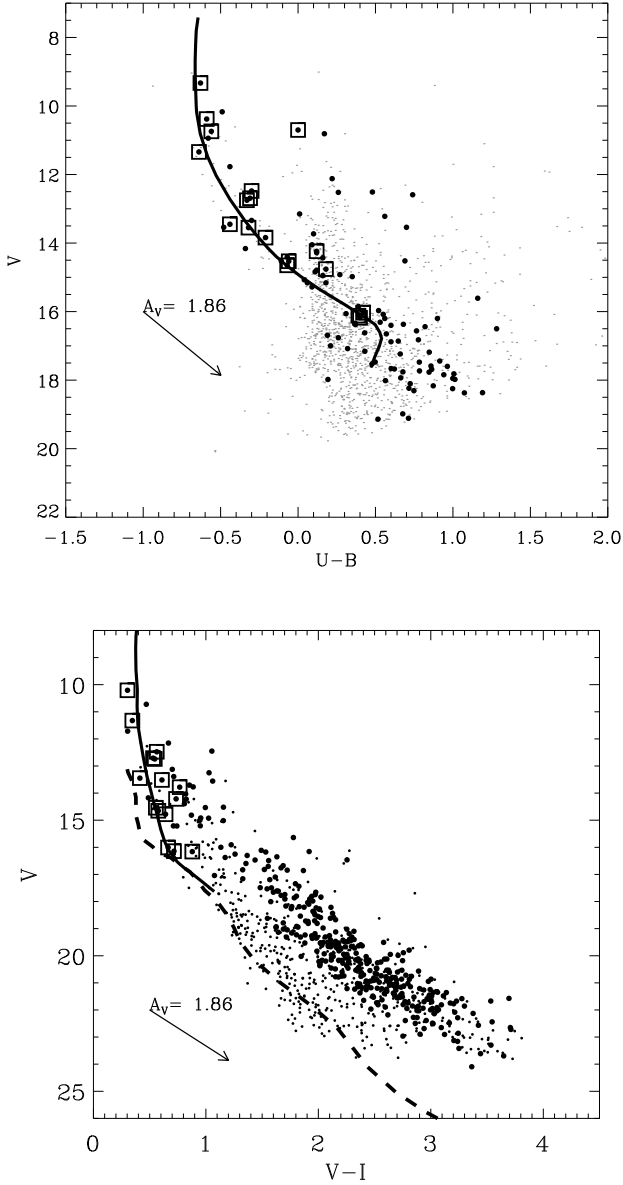}
      \caption{V vs. U-B diagram  from the \citet{mass95} and \citet{shar07} photometry (upper panels) and 
  V vs. V-I diagram of all the objects with errors in V-I smaller than 0.1 and
  located in the regions with A$_{\rm V}>3.8$ (bottom panels).
  Filled circles are the diskless candidate   cluster members; squares are those with 0.5$<$E(B-V)$<$0.7; 
  solid line is the  3.0\,Myr isochrone of solar metallicity using   E(B-V)=0.6 while
  dashed line is the solar metallicity ZAMS of \citet{sies00} using E(B-V)=0.4;  
  both are drawn assuming a distance of 3.6\,kpc. 
  The reddening vector obtained using the \citet{muna96} reddening laws is also shown.}
	  \label{distancefig}
    \end{figure}

  However, in the V vs. U-B diagram, the MS population is quite well separated from the
  bulk of the remaining objects that are field stars and candidate PMS cluster members.
This is evident in the upper panels of Fig.\,\ref{distancefig}, obtained 
by using the UBV photometry by \citet{mass95} and \citet{shar07}, and where diskless 
candidate cluster members selected by us are
  indicated with filled circles.  Among these objects, we indicate with squares those that according to the U-B vs. B-V 
diagram have 0.5$<$E(B-V)$<$0.7. We use these objects here to 
   constrain the cluster distance by comparing again the 
  solar metallicity isochrone  of 1.5\,Myr  of \citet{mari08}
  reddened with E(B-V)=0.6.
  By taking advantage of the slightly curved shape of the isochrone in this CMD, 
we find that a good fitting
  is achieved if a distance of 3.6\,kpc is adopted with an error of about 200\,pc. With our data, we can 
rule out for NGC\,1893 a  distance of 3200\,pc or 4300\,pc that are values previously given in the literature \citep{shar07,tapi91,mass95}.

  A  distance of 3.6\,kpc for NGC\,1893 is  also consistent with that derived  by using an independent 
method 
based on the nebula properties. In fact, as already done in previous works \citep{pris05,guar07}, we assume that
 the nebula surrounding the cluster
    obscures many  background objects and allows us to see mainly foreground stars, that are
  mostly in MS, with distance smaller or equal to the nebula distance. Among these latter, 
  those located at the nebula distance
  define the blue envelope of the CMD diagram that can be fitted by the ZAMS locus. In the plausible
 assumption that the nebula is located at the same distance of  NGC\,1893 and by considering only very high 
reddened regions around the cluster (to minimize the spread due to background objects), it is possible to derive the cluster
  distance by  fitting the ZAMS to the blue envelope of the CMD.

  By a visual inspection of our observations in the optical, JHK and Spitzer/IRAC bands, we note that 
  the nebula is not evenly spatially distributed, lying mostly in the western region of our FoV. For this reason,
  we selected the MS stars to be used for the ZAMS isochrone fitting by using the most obscured regions.
 The latter
  have been selected by means of a reddening map based on the distribution of the H-K colors, as done in 
\citet[][see this paper for details]{dami06}  by using H-K=0.1 as limit for the intrinsic color expected for background giants.
  As in \citet{dami06},  for the map computation we excluded all objects classified as candidate cluster members. 
  The  A$_{\rm V}$ map has been derived by using a cell size of 3.1\min$\times$ 2.6\min~ within the FoV defined
 by our NICS/JHK observations and
 by  the E(K-H) map, obtained as the median of H-K colors in each cell minus the expected 0.1 value.
 We find that the whole region is affected by $2.2\lesssim {\rm A}_{\rm V}\lesssim4$, with the reddest regions being the western ones. As already mentioned, the whole region is not affected by very 
high A$_{\rm V}$ values and this can be also confirmed  by the distribution of field stars in the J-H vs. H-K diagrams. This 
is likely due to a low
  absolute reddening affecting the galactic anti-center direction.

  By using the resulting  A$_{\rm V}$ map, we selected all the objects within the regions with A$_{\rm V}>3.8$
  and error in the V-I colors smaller than 0.1; the  CMD diagram  of these objects (indicated with dots)
 is shown in the bottom panels of Fig.\,\ref{distancefig}, where filled circles are diskless candidate  members. 
The dashed curve in the figure represents the solar metallicity  ZAMS of \citet{sies00} shifted at the distance
 d=3.6\,kpc and  using E(B-V)=0.4, that is the minimum reddening we associate to field stars according to the bright stars
 in this diagram, while the solid line is the solar metallicity 1.5\,Myr isochrone of \citet{mari08} 
 for   stars with mass M$>$1.5M$_\odot$,
  adopting the same distance and E(B-V)=0.6,
  that is the mean cluster reddening we derived in the previous section.

  As already mentioned, we derive the cluster distance as that value for which a good ZAMS fitting to MS stars of the blue envelope of this diagram  is achieved. We note that, since the maximum
 A$_{\rm V}$ is about 4,
  even in the most obscured regions, a little fraction of faint background objects could be visible. In fact,
  as shown in Fig.\ref{distancefig} (bottom panel), objects with V$>21$ do not follow the ZAMS shape, likely because this region is populated by field stars or 
extragalactic objects, that are expected to be visible in 
the galactic anti-center direction where the cluster NGC\,1893 is located.  Therefore, for the ZAMS fitting we consider only field stars with V$<21$; nevertheless,
  since the bright population of field stars (V$\lesssim 17$) is poorly populated, we considered, in this range,
 the candidate  cluster members with $V\lesssim 16$ that are on the MS phase and with 0.5$<$E(B-V)$<$0.7, indicated 
with squares.
  We note that V$\simeq16$ is the magnitude of the Turn On, i.e. the magnitude where the PMS joins the cluster MS.
  We fitted the candidate MS cluster members assuming the mean cluster reddening E(B-V)=0.6 and we 
 found that 
  a good fit is  obtained both for the MS stars in the blue envelope and for the 
  bright candidate cluster members on MS if a distance of $\sim$3.6\,kpc (see bottom panel of Fig.\ref{distancefig}) is used. 
By taking the solar galactocentric distance of 8.5\,Kpc, we deduce that NGC\,1893 is located at a distance of about  12.1\,Kpc
from the Galactic Center. 

  \section{Comparison with literature cluster parameters\label{compareparam}}
  The cluster distance we  derived (3.6$\pm$0.2\,kpc) is marginally consistent
 with that recently derived by \citet{shar07} (d=3250\,pc) by fitting the 4\,Myr 
isochrone of \citet{bert94} to the MS field stars and assuming E(B-V)=0.4, 
that is the value  taken for the MS field stars. This method is similar to 
that used by us, since we note that
  the  4\,Myr isochrone of \citet{bert94} corresponds rather to the ZAMS locus, if  we consider that stars
  4\,Myr old  with masses smaller than about 2\,M$_\odot$ are yet in PMS. 

  We  find a good agreement with the value derived by \citet{cuff73} who, using UBV magnitudes,
   derived that the reddening in front of 
  the cluster amount to E(B-V)=0.4, while the distance is 3.6\,kpc. Among  the first distance values published 
  for NGC\,1893,  \citet{beck71} found d=3700\,pc and  
  A$_{\rm V}=1.68$, corresponding to  E(B-V)=0.54, by using simultaneously two color-magnitude diagrams.
  \citet{moff72}  derived 
a distance of 3980\,pc and a reddening E(B-V)=0.55 using photographic UBV photometry
  and the ZAMS fitting to the unreddened CMD.
  \citet{hump78} found a distance modulus of 12.52 corresponding to
  a distance of about 3200\,pc, by using V, B-V photometry of three O type main sequence stars of NGC\,1893 
  in  a paper devoted to study fundamental properties of the most luminous stars in our Galaxy.

  A larger distance of 4300\,pc  has been derived by \citet{tapi91} by using Str$\ddot{\rm{o}}$mgren $ubvy$, H$\beta$ and JHK photometry. 
  They derived spectral types and approximate luminosity classes from the $m_1$ and $c_1$ indices that together with literature slit spectral types are used to estimate reddening and absorption and therefore intrinsic colors and magnitudes for candidate cluster members. The corrected distance modulus is derived as the difference between the unabsorbed V magnitude and the absolute magnitude predicted for  the given spectral type. A similar value, equal to 4800\,pc, has been 
  found by \citet{fitz93}, by using again Str$\ddot{\rm{o}}$mgren $ubvy$ photometry and the theoretical ZAMS fitting.

  CCD UBV photometry and multiobject fiber spectroscopy are used by \citet{mass95} to determine  distance 
  and reddening equal to 4400\,pc and E(B-V)=0.53$\pm$0.2, respectively. These values are derived by using individual reddening
  from spectroscopy and appropriate spectral type-M$_{\rm V}$ calibration for O and B-type stars.  The derived distance
  has been obtained by including only stars with inferred color excess E(B-V) between 0.41 and 0.70 

\citet{vall99} use J and K photometry and UBV literature photometry to derive E(J-K) from 0.15 to 0.5, corresponding
to 0.25$<$E(B-V)$<$0.8 in the whole region they study; in addition they find two main dark regions where E(J-K) is 0.45-0.55
(or E(B-V)=0.7-0.9) that is higher than the value derived in the regions outside the dark clumps.

The most discordant distance value for NGC\,1893 with respect to that found in this paper
 is that derived by \citet{marc01} equal to 6000\,pc; they
 use  photometric Str$\ddot{\rm{o}}$mgren indices (magnitude limit V$\simeq$15.9)
 to compute the interstellar reddening $E(b-y)$  of cluster members.
They find values up to $E(b-y)=$0.607 but they estimate the average interstellar reddening
$E(b-y)=0.33\pm0.03$
by considering only stars with $E(b-y)<0.4$. They use this average reddening
to derive individual magnitudes and colors ($V_0$, $(b-y)_0$ and $c_0$).
By fitting an empirical ZAMS in the M$_{\rm V}$ vs. $c_0$ diagram they derive 
a dereddened distance modulus of 13.9. Considering  that the  cluster is affected
by a significant differential reddening, we suppose that the photometric intrinsic
parameters derived by \citet{marc01}, based on the average reddening  could be the
origin of their overestimate of the cluster distance.  

  \section{Masses and Ages}
Assuming the cluster parameters derived by us in the previous sections, we 
 was able to 
compute masses and ages of all objects we selected as candidate cluster members. However,
as we mentioned before, the position in the color-magnitude diagram of class\,II candidate members
can be influenced by effects due to the presence of circumstellar disk. Since our data
do not allow to disentangle the purely photospheric spectrum of the stars, to which
theoretical models are referred, from that in excess
due to such effects, we assume that masses and ages of diskless  stars
can be accurately derived within the uncertainties due to photometric errors, reddening and to the theoretical models,
 while masses and ages of class\,II  stars should be taken with caution, 
especially those of the 170 candidate members falling outside  the
PMS region with ages apparently older than 10\,Myr.

\begin{figure*}
    \centering
    \includegraphics[width=\textwidth]{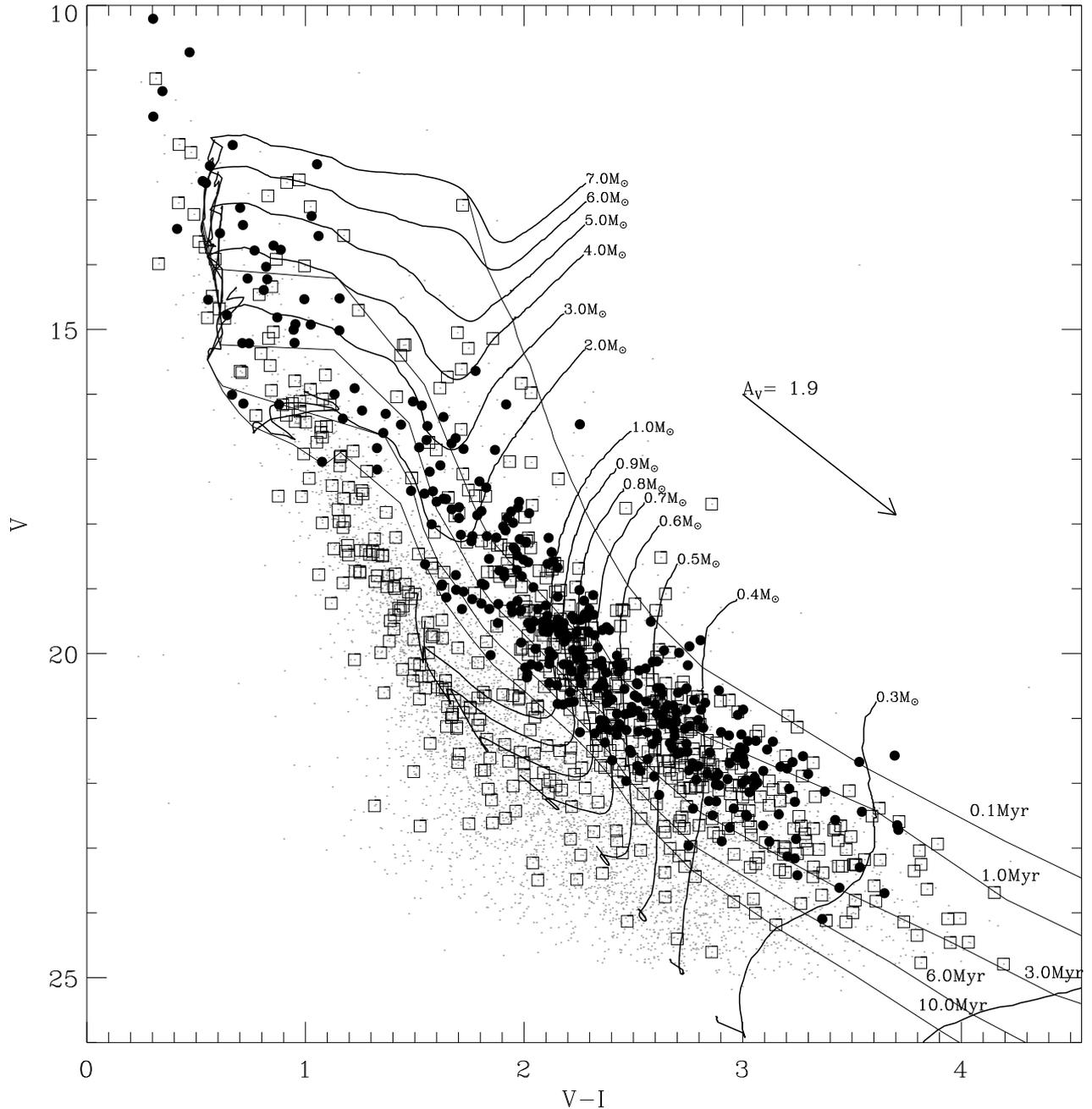}
      \caption{V vs. V-I diagram of all objects with optical magnitudes in our catalog (dots).
diskless  and class\,II candidate  members are indicated with filled circles and empty squares, respectively.
Lines are the \citet{sies00} tracks and isochrones
 of masses and ages indicated on each line.}
	  \label{viv_models}
    \end{figure*}

Masses and ages are computed by interpolating  the solar metallicity 
\citet{sies00} tracks
on the V vs. V-I plane, using the  TRIGRID idl function. Figure\,\ref{viv_models} shows the
V vs. V-I diagram with the \citet{sies00} tracks and isochrones superposed.

\begin{figure}
    \centering
    \includegraphics[width=9cm]{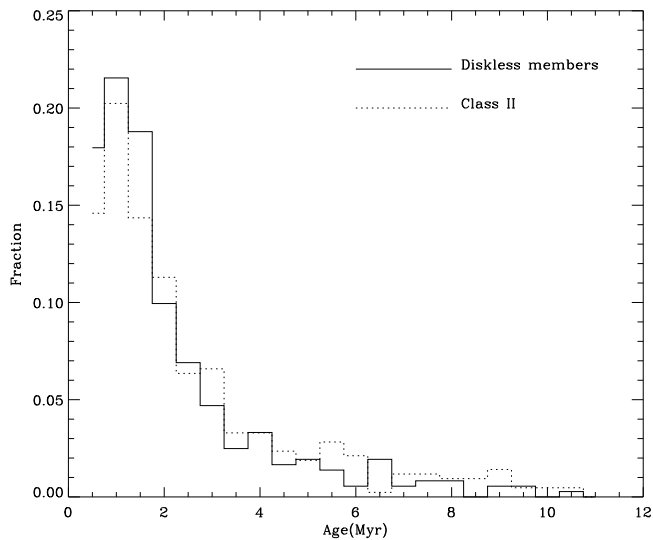}
      \caption{Age distribution of diskless  and class\,II candidate members in the PMS region. Class\,II YSOs
 with anomalous blue colors are not included in this sample.}
	  \label{mass_age_statistics}
    \end{figure}

For the diskless  sample we find stars with masses between 0.3 and 
6.7\,M$_\odot$\footnote{the upper limit is imposed by the adopted \citet{sies00} models.}; 
we have further 11 objects
classified as diskless candidate members with  mass larger than 7\,M$_\odot$; 
the median value and the  standard deviation  of diskless member ages are 1.4\,Myr and 1.8\,Myr,
respectively.
For the class\,II sample, if we discard the 170 stars with optical anomalous photometry,
  we find stars with masses between 0.2 and 6.9\,M$_\odot$; we have further 10 objects
classified as class\,II candidate members with  mass larger than 7\,M$_\odot$; 
the median value and the  standard deviation  
of ages of candidate  members with disk are 1.6\,Myr and 2.2\,Myr, respectively, with again few stars being 9-10\,Myr old.
We conclude that, apart from the 170 objects with anomalous optical indices, the remaining class\,II stars
show mass and age distributions very similar to those of diskless  YSOs as it is shown in Fig.\,\ref{mass_age_statistics} where the age distribution of diskless  and class\,II candidate members are shown.
This reflects the similar spread observed in the V vs. V-I diagram by the two populations of candidate
members. Apparently the colors of class\,II candidate members are not  substantially modified by phenomena
related to accretion.
 The peak in the observed mass function is among 0.4 and 0.5\,$M_\odot$; however the
 IMF will be computed in an other work where we will discuss the biases due to the adopted methods and where
we will compare the IMF with that of other clusters of similar age.


  \section{Summary and conclusions}
We used new deep optical NIR data in the VRIJHK and H$_\alpha$ bands together with literature
X-ray and Spitzer-IRAC data to compile a multiband photometric catalog in the direction of  the
young distant open cluster NGC\,1893, located  towards the galactic anti-center. 
Using these data and different membership criteria, we made the most complete census of candidate   members
available in the literature for this cluster. Reddening-independent indices, involving optical and/or NIR
colors, allowed us to derive a list of 1061 candidate members with a circumstellar disk, with 125 among them being
also H$_\alpha$ emitters; among the disk bearing
candidate members, 170 show anomalous optical magnitudes and colors that are likely
due to effects of gas and/or dust in the disk. In addition, X-ray detections and optical
magnitudes, allowed us to distinguish 415 candidate members without disk. 

We used these last 415 YSOs and their photometric properties, to assess the cluster parameters,
viz. interstellar reddening and distance. Disk-less candidate  members  are, in fact, the most reliable objects
to be compared with theoretical tracks and isochrones, since their colors are purely photospheric
and do not suffer from additional effects due to the presence of circumstellar disks.

Bright diskless  members in the U-B vs. B-V diagram, obtained from literature data, are consistent with
an interstellar reddening E(B-V)=0.6$\pm$0.1. This value is also consistent using the same bright
candidate members in the independent 
V vs. V-I diagram and most of the low mass diskless candidate  members in the V-I vs. J-K diagram.
 However, we find evidence of differential reddening in this region.

Using  the V vs. B-V (from the literature)  diagram,
and assuming the mean cluster reddening E(B-V)=0.6, we find that the main sequence members
selected in this work are well fitted by a  1.5\,Myr isochrone, if a distance of 3.6$\pm$0.2\,kpc
is adopted. We note that in the MS cluster region,  isochrones degenerate and therefore
our distance estimate is independent  of the age adopted for MS members. The value of 3.6\,kpc
is consistent with the cluster distance we derive by using  the main sequence blue envelope 
traced in the V vs. V-I diagram  by foreground field stars located
at the cluster distance. Therefore, using independent methods and data, we are able to firmly 
evaluate the cluster distance for which values between 3250 and 6000\,pc are given in the most recent
literature.

Finally, we derived masses and ages of selected candidate members in the optical PMS region down to 
about 0.2\,M$_\odot$; median ages for diskless  and class II stars, are, respectively,
 1.4 and 1.6\,Myr, with a standard deviation of 1.8 and 2.2\,Myr for the two samples.
 Few cluster candidate members are older with ages up to about 10\,Myr. 

If we consider that 1001 of the 1061 class\,II candidate members are within the Chandra FoV, where we have
selected the diskless candidate  members, we find a disk fraction of about 71\% of stars. This value
is an upper limit since the diskless member selection is based on the X-ray observations that reach
a mass limit higher than that we can reach with NIR observations used to select class\,II stars.
However, the disk fraction we find is similar to the 67\% value we found in Paper I and in agreement
with the disk fraction found in clusters of similar age \citep{hais01}, if we assume for NGC\,1893 a median 
age of 1.5\,Myr. A detailed analysis of the disk fraction as a function of stellar masses and cluster location 
will be the subject of a forthcoming  paper (Sanz-Forcada et al. 2010, in preparation).

We therefore conclude that, despite its peculiar location in the Galaxy,
 NGC\,1893 includes a rich population of young stars, with general 
properties similar
to those found in young clusters in the solar   neighborhood.
\begin{acknowledgements}
  Part of this work was financially supported by the  PRIN-INAF (P.I. Lanza)
  and the EC MC RTN CONSTELLATION (MRTN–CT–2006–035890). 
     We thank Donata Randazzo for carefull reading of the paper.
\end{acknowledgements}

 \bibliographystyle{aa} 
  \bibliography{report}

  \addtocounter{table}{1}
\Online
\begin{appendix}
\section{Photometric calibration}
\subsection{DOLORES}
Photometric calibration of DOLORES observations  was computed by using
the images of the \citet{land92} standard field SA\,98. We  used the
{\em v, r, i, h$_\alpha$} instrumental magnitudes, and  considered the V,
R and I 
magnitudes of the Johnson-Kron-Cousins photometric system of standard
stars in these fields \citep{ste00}; H$_\alpha$ is the magnitude in the
corresponding non-standard filter.  
We   calculated the transformation coefficients
between the instrumental and standard systems by using the CCDSTD code \citep{stet05}  
and the following equations:
\begin{eqnarray}
\label{calib_eq}
v & = & V + A_0 + A_1 \, Q + A_2 \, (V-I) + A_3 \, (V-I)^2 + A_4 \, X\, Y + \nonumber \\
  &   & + A_5 \, X + A_6 \, Y + A_7\,  X^2 + A_8\,  Y^2  \nonumber \\
r & = & R + B_0 + B_1 \, Q + B_2 \, (R-I) + B_3 \, (R-I)^2 + B_4 \, X\, Y + \nonumber \\
  &   & + B_5\,  X + B_6\,  Y + B_7\,  X^2 + B_8\,  Y^2  \\
i & = & I + C_0 + C_1 \, Q + C_2 \, (V-I) + C_3\,  (V-I)^2 + C_4 \, X\, Y + \nonumber \\
  &   & + C_5\,  X + C_6\,  Y + C_7\,  X^2 + C_8\,  Y^2  \nonumber \\
h_\alpha & = & H_\alpha + D_0 + D_1\, Q + D_2\,  (R-I) + D_3\,  (R-I)^2 + D_4\,  X\, Y + \nonumber \\
  &   & + D_5\,  X + D_6\,  Y + D_7\,  X^2 + D_8\,  Y^2  \nonumber
\end{eqnarray}
where $Q$ is the airmass, $A_0$, $B_0$ and $C_0$ are the magnitude
zero points; $A_1$, $B_1$ and $C_1$ are 
the extinction coefficients;
$A_2$, $B_2$ and $C_2$ and $A_3$, $B_3$ and $C_3$  are the color terms, and the rest of terms are
related to the geometrical coordinates ($X, Y$) of the
plate. Stars with magnitude residual values above the 3$\sigma$ level
were not considered for the coefficient fitting. 
Coefficients calculated during calibration are listed in
Table\,\ref{coefficients}. DOLORES extinction terms are set to those
provided by the telescope web site. We
 followed the same nomenclature for the H$\alpha$ filter, but since we have
no standard measurements in this band, we  applied the same calibration as
for the R band (therefore $D_i=C_i$) in the case of DOLORES. 
\begin{table*}
\tabcolsep 3.4pt
\caption{Coefficients of the transformations to the standard system, given in  the
Eq.s\,\ref{calib_eq}, for
  each filter and night for DOLORES (D)$^a$ and CAFOS (C)}
\label{coefficients}
\begin{tabular}{lcccccccccccc}
  \hline \hline
Instr. & Date & Filter & Coeff. & \multicolumn{9}{c}{ $A_i$, $B_i$, $C_i$, $D_i$ coefficient indices }  \\
& &  &  & 0 & 1 & 2 & 3 & 4 & 5 & 6 & 7 & 8 \\
  \hline
D & 2007/09/21 & V & $A_i$ &     -0.874     & 0.15 		&  0.087      & \ldots 	   & \ldots 	      & -0.068    &  0.138    & \ldots                & -0.045  \\
  &            &   &       &     $\pm$0.014 &     		&  $\pm$0.007 & \ldots 	   & \ldots 	      &$\pm$0.006 &$\pm$0.021 & \ldots                &$\pm$0.010 \\
D &          "  & R & $B_i$ &    -0.823     & 0.11 		& -0.002     & \ldots 	   & \ldots	      & -0.072      & -0.007       & \ldots           & \ldots 	  \\   
   &          "  &   &       &   $\pm$0.008  &    		& $\pm$0.009 & \ldots 	   & \ldots	      & $\pm$0.004  &   $\pm$0.003 & \ldots           & \ldots 	  \\    
 D &          "  & I & $C_i$ &    -0.531     & 0.07 		& -0.035     & \ldots 	   & \ldots 	      &  0.420      &  0.393       & -0.191           & -0.181 \\
   &          "  &   &       &    $\pm$0.020 &  		& $\pm$0.006 & \ldots 	   & \ldots 	      &  $\pm$0.026 &  $\pm$0.017  & $\pm$0.010       & $\pm$0.008 \\
 D &  2007/10/18 & V & $A_i$ &    -1.039     & 0.15 		&  0.103     & \ldots 	   & \ldots 	      & -0.007	    &  0.015       & \ldots           & -0.006\\
   &      	 &   &       &    $\pm$0.004 &   		& $\pm$0.002 & \ldots 	   & \ldots 	      & $\pm$0.002  & $\pm$0.007   & \ldots           &$\pm$0.003 \\ 
 D &          "  & R & $B_i$ &    -1.100     & 0.11 		&  0.015     & \ldots 	   & \ldots 	      &  0.013	    &  0.011	   & \ldots           & \ldots 	  \\
   &          "  &   &       &    $\pm$0.003 & 			& $\pm$0.004 & \ldots 	   & \ldots 	      & $\pm$0.002  &  $\pm$0.002  & \ldots           & \ldots 	  \\
 D &          "  & I & $C_i$ &    -0.572     & 0.07		&  0.018     & \ldots 	   & \ldots 	      &  0.237      &  0.224       & -0.081           & -0.117   \\
   &          "  &   &       &    $\pm$0.009 &  		& $\pm$0.004 & \ldots 	   & \ldots 	      & $\pm$0.019  &  $\pm$0.010  & $\pm$0.011       & $\pm$0.005 \\ 
 D &  2007/11/14 & V & $A_i$ &    -1.019     & 0.15 		&  0.090     & \ldots 	   & \ldots 	      & -0.014      &  0.072       & \ldots           & -0.026  \\  
   &             &   &       &    $\pm$0.013 &  		& $\pm$0.004 & \ldots 	   & \ldots 	      & $\pm$0.016  &  $\pm$0.011  & \ldots           & $\pm$0.005 \\  
 D &          "  & R & $B_i$ &    -1.050     & 0.11 		& -0.007     & \ldots	   & \ldots 	      &  0.043      &  0.002       & -0.026           & \ldots 	  \\
   &          "  &   &       &    $\pm$0.005 &   		& $\pm$0.004 & \ldots	   & \ldots 	      & $\pm$0.008  & $\pm$0.002   & $\pm$0.003       & \ldots 	  \\ 
 D &          "  & I & $C_i$ &     -0.537    & 0.07 		& -0.019     & \ldots 	   & \ldots 	      &  0.307      &  0.239       & -0.136           & -0.128  \\
   &          "  &   &       &    $\pm$0.010 &  		& $\pm$0.004 & \ldots 	   & \ldots 	      &  $\pm$0.014 &  $\pm$0.010  & $\pm$0.006       & $\pm$0.005 \\
 C & 2007/10/11 & V & $A_i$ &     2.482      & 0.118  		& -0.022     &  0.015      & -0.026           & -0.236      & -0.127       &  0.139           &  0.086  \\
   &            &   &       &     $\pm$0.019 & $\pm$0.004 	& $\pm$0.015 & $\pm$0.005  & $\pm$0.011       & $\pm$0.022  & $\pm$0.023   &  $\pm$0.010      &  $\pm$0.012 \\ 
 C &         "  & R & $B_i$ &     1.820      & 0.045      	&  0.030     & -0.056      & -0.147           &  0.261      &  0.151       & \ldots           & \ldots 	 \\
   &         "  &   &       &     $\pm$0.042 & $\pm$0.011 	& $\pm$0.049 & $\pm$0.031  & $\pm$0.061       & $\pm$0.082  &  $\pm$0.091  & \ldots           & \ldots 	 \\ 
 C &         "  & I & $C_i$ &    2.467       & 0.051     	& -0.082     & -0.015      & -0.039           &  0.206      &  0.344       & -0.132           & -0.208  \\
   &         "  &   &       &    $\pm$0.021  & $\pm$0.005 	&$\pm$0.009  & $\pm$0.004  & $\pm$0.025       &  $\pm$0.030 &   $\pm$0.040 & $\pm$0.021       & $\pm$0.022 \\
 C &  " &H$_\alpha$ & $D_i$ &    -1.344      & 1.088     	&  0.340     & \ldots      &  0.063           & -0.333      & -0.384       &  0.205           & 0.238 \\
   &  " &           &       &     $\pm$0.027 &$\pm$0.323 	& $\pm$0.036 & \ldots      &  $\pm$0.014      & $\pm$0.028  & $\pm$0.029   &  $\pm$0.012      &$\pm$0.016 \\
 C& 2008/01/05 & V & $A_i$  &    2.642       & \ldots          	&  0.063     & -0.012      & -0.018           &  0.190      &  0.118       & -0.094           & -0.098  \\
  &            &   &        &     $\pm$0.012 & \ldots          	& $\pm$0.011 & $\pm$0.004  & $\pm$0.008       &  $\pm$0.014 & $\pm$0.013   & $\pm$0.007       &$\pm$0.007 \\
 C&         "  & R & $B_i$ &     1.940       & \ldots          	&  0.038     & -0.037      & -0.010           &  0.201      &  0.129       & -0.099           & -0.131  \\
  &         "  &   &       &      $\pm$0.010 & \ldots          	& $\pm$0.014 & $\pm$0.008  & $\pm$0.007       &  $\pm$0.012 & $\pm$0.011   & $\pm$0.005       & $\pm$0.006 \\
 C&         "  & I & $C_i$ &     4.497       & \ldots          	& -0.108     & -0.006      & -0.222           & -0.200      &  0.233       &  0.174           & -0.043 \\
  &         "  &   &       &     $\pm$0.021  & \ldots          	& $\pm$0.011 &$\pm$0.003   & $\pm$0.015       & $\pm$0.027  & $\pm$0.024   & $\pm$0.013       & $\pm$0.013 \\
 C&  " &H$_\alpha$ & $D_i$ &     2.948       & \ldots          	& -0.373     &  0.374      &  0.027           & -0.266      & -0.594       &  0.162           &  0.347  \\
  &  " &           &       &     $\pm$0.105  & \ldots          	& $\pm$0.154 & $\pm$0.127  & $\pm$0.060       & $\pm$0.130  &$\pm$0.100    & $\pm$0.052       &$\pm$0.059 \\ 
 C&2008/01/09&H$_\alpha$& $D_i$&1.724        & \ldots    	&  0.045     & -0.013      &  0.118           &  0.322      & -0.266       & -0.176           &  0.028 \\
  &          &          &      &$\pm$0.045   & \ldots    	& $\pm$0.051 &$\pm$0.035   &$\pm$0.026        & $\pm$0.049  & $\pm$0.050   &$\pm$0.019        &$\pm$0.024 \\
\hline
\multicolumn{5}{l}{ \scriptsize{
 {\it Notes:} $^a$DOLORES observations in H$\alpha$ have same coefficients as for the R filter}} & \multicolumn{6}{c}{}\\

\end{tabular}
\end{table*}

\subsection{CAFOS}

We tried to calibrate the CAFOS observations following the same procedure
as for DOLORES, using the site extinction terms as provided by
\citet{sanchez07}. However we noticed substantial discrepancies in the
final magnitudes calculated for CAFOS when compared to DOLORES
results. The variable atmospheric conditions detected during the
observations, also noted in the seeing, could be responsible of the
inconsistencies. We decided to use the DOLORES
photometry to calibrate the CAFOS fields: we selected the local
standards used in DOLORES fields that are also common to the CAFOS
FoV, and used a total of 417 CAFOS counterparts as standards for the
calibration, so we  calibrated the CAFOS observations in the
same photometric system of DOLORES. The second and third nights of CAFOS
observations also required to set the ``cloud'' variable in the CCDSTD
setting. As explained by \citet{ste00} this
variable allows
to consider different atmospheric conditions (thin clouds) during the
night, setting 
all the extinction terms to 0 in this case (Table\,\ref{coefficients}),
and fitting the zero points and the other terms.      
CAFOS observations used the DOLORES standard stars falling in the field to
calibrate the H$\alpha$ observations, with $D_i$ coefficients as
listed in Table~\ref{coefficients}.

  \subsection{NICS}
   We used the standalone DAOPHOT II/ALLSTAR code 
\citep{stet87} to obtain the instrumental photometry for the
  combined images obtained with SNAP for each filter (JHK) and for each of the 16 fields
  observed around the cluster NGC\,1893. For most of the images we performed the PSF photometry using
  the Moffat function with $\beta=2.5$, typically used to model stellar profiles as an analytic first
approximation to the PSF, while for 6 images
  we used the more complex Penny function in order to take into account the elongated shape of the
  PSF likely due to some aberration. In addition we treated the variable PSF  by using
 the DAOPHOT option that considers a PSF which varies quadratically with position in the frame.  

Since the PSF photometry is relative to the model stellar profile of a given frame,
we  needed to derive the aperture correction to the instrumental photometry; to this aim 
we performed aperture photometry
  at different radii  on a sample of isolated and relatively bright stars from which we 
derived growth curves with  the DAOGROW code \citep{stet90}. The aperture correction has been computed using the selected stars  as the median of the
  difference between the PSF magnitudes and the aperture magnitudes obtained at the radius including all
  the stellar flux.

  In order to discard false identifications due to the spots either 
of very bright objects or just falling at the 
  edge of the images, we selected the list of objects detected
 by considering only those following the
  typical exponential profile of the magnitude errors. Then for each of the 16 observed fields,
  we merged with DAOMATCH/DAOMASTER \citep{stet87} the three lists with JHK magnitudes in order to 
  have a single list with objects having at least two among the JHK magnitudes.

  The NICS filters used for our observations (Js, H, K') are those of the Mauna
Kea Observatories (MKO) near-infrared filter set \citep{ghin02} and therefore we
calibrate our catalog in the MKO photometric system. To this aim we use as standard
stars, the 2MASS counterparts falling in our fields with PH\_QUAL flag equal to 'AAA' and
 CC\_FLG flag equal to  '000'. We first converted the 2MASS 
catalog in the MKO system by using the inverted transformations given in the 2MASS 
web page\footnote{http://www.ipac.caltech.edu/2mass/releases/allsky/doc/sec6\_4b.html}
maintened by J. Carpenter \citep[see also][]{legg06}.

By using instrumental magnitudes for the standard stars, that we indicate as j, h, k and those of the 2MASS catalog in the MKO system,
that we indicate as J$_{\rm MKO}$, H$_{\rm MKO}$, K$_{\rm MKO}$
we performed for each field a linear fit of the magnitude differences as a function of the colors as
in the following equations:
\begin{center}
\begin{eqnarray}\label{equatnics}
%
{\rm (j-J_{MKO})=A_0+A_1(j-h)}\nonumber \\ 
{\rm (j-J_{MKO})=B_0+B_1(j-k)}\nonumber \\ 
{\rm (h-H_{MKO})=C_0+C_1(j-h)}  \\ 
{\rm (h-H_{MKO})=D_0+D_1(h-k)}\nonumber \\  
{\rm (k-K_{MKO})=E_0+E_1(j-k)}\nonumber \\ 
{\rm (k-K_{MKO})=F_0+F_1(h-k)}.\nonumber  
\end{eqnarray}
\end{center}
From the linear fit we derived the zero points, indicated by coefficients with subscript 0
and the color terms, indicated by coefficients with subscript 1 given  in Table\,\ref{nicscoefftab}.
\begin{table*}
\caption{Zero points (coefficients with subscript 0) and color terms (coefficients with subscript
 1) derived from the linear fit of equations (\ref{equatnics}) to transform instrumental
NICS magnitudes to the MKO system.}
\label{nicscoefftab}
\begin{tabular}{ccccccccccccc}
  \hline \hline
f & A$_0$ & A$_1$ & B$_0$ & B$_1$ & C$_0$ & C$_1$ & D$_0$ & D$_1$ & E$_0$ & E$_1$ & F$_0$ & F$_1$ \\  
\hline
   1  &  -22.505    &     0.039     &  -22.487    &     0.027     &  -22.522    &    -0.056     &  -22.598    &    -0.108     &  -21.957    &    -0.007     &  -21.964    &    -0.015    \\
  &$\pm$    0.013 &$\pm$     0.018&$\pm$    0.007 &$\pm$     0.013&$\pm$    0.015 &$\pm$     0.019&$\pm$    0.015 &$\pm$     0.043&$\pm$    0.009 &$\pm$     0.014&$\pm$    0.015 &$\pm$     0.044    \\
   2  &  -22.577    &     0.064     &  -22.548    &     0.049     &  -22.740    &     0.014     &  -22.673    &     0.142     &  -22.178    &     0.072     &  -22.077    &     0.205    \\
  &$\pm$    0.026 &$\pm$     0.040&$\pm$    0.010 &$\pm$     0.028&$\pm$    0.029 &$\pm$     0.044&$\pm$    0.042 &$\pm$     0.099&$\pm$    0.013 &$\pm$     0.034&$\pm$    0.047 &$\pm$     0.114    \\
   3  &  -22.388    &    -0.014     &  -22.397    &    -0.004     &  -22.581    &    -0.039     &  -22.617    &    -0.020     &  -21.928    &    -0.010     &  -21.942    &    -0.024    \\
  &$\pm$    0.025 &$\pm$     0.036&$\pm$    0.009 &$\pm$     0.026&$\pm$    0.029 &$\pm$     0.041&$\pm$    0.048 &$\pm$     0.098&$\pm$    0.010 &$\pm$     0.029&$\pm$    0.043 &$\pm$     0.089    \\
   4  &  -22.410    &    -0.107     &  -22.459    &    -0.076     &  -22.519    &    -0.231     &  -22.757    &    -0.255     &  -22.079    &    -0.030     &  -22.148    &    -0.157    \\
  &$\pm$    0.016 &$\pm$     0.028&$\pm$    0.006 &$\pm$     0.022&$\pm$    0.020 &$\pm$     0.032&$\pm$    0.034 &$\pm$     0.082&$\pm$    0.008 &$\pm$     0.025&$\pm$    0.035 &$\pm$     0.083    \\
   5  &  -22.504    &     0.044     &  -22.485    &     0.034     &  -22.609    &    -0.101     &  -22.708    &    -0.100     &  -22.116    &     0.014     &  -22.128    &    -0.041    \\
  &$\pm$    0.012 &$\pm$     0.019&$\pm$    0.005 &$\pm$     0.015&$\pm$    0.013 &$\pm$     0.020&$\pm$    0.023 &$\pm$     0.058&$\pm$    0.007 &$\pm$     0.017&$\pm$    0.025 &$\pm$     0.063    \\
   6  &  -22.471    &     0.053     &  -22.450    &     0.048     &  -22.581    &    -0.053     &  -22.583    &     0.078     &  -22.081    &     0.056     &  -21.998    &     0.189    \\
  &$\pm$    0.014 &$\pm$     0.022&$\pm$    0.006 &$\pm$     0.017&$\pm$    0.015 &$\pm$     0.025&$\pm$    0.023 &$\pm$     0.060&$\pm$    0.007 &$\pm$     0.017&$\pm$    0.021 &$\pm$     0.056    \\
   7  &  -22.063    &    -0.023     &  -22.073    &     0.011     &  -22.074    &    -0.069     &  -22.076    &     0.094     &  -21.682    &    -0.004     &  -21.683    &    -0.001    \\
  &$\pm$    0.010 &$\pm$     0.024&$\pm$    0.005 &$\pm$     0.018&$\pm$    0.013 &$\pm$     0.029&$\pm$    0.015 &$\pm$     0.052&$\pm$    0.006 &$\pm$     0.019&$\pm$    0.014 &$\pm$     0.049    \\
   8  &  -22.627    &     0.257     &  -22.493    &     0.198     &  -22.670    &    -0.067     &  -22.664    &     0.114     &  -22.145    &     0.246     &  -22.510    &    -0.891    \\
  &$\pm$    0.025 &$\pm$     0.035&$\pm$    0.011 &$\pm$     0.037&$\pm$    0.029 &$\pm$     0.037&$\pm$    0.051 &$\pm$     0.105&$\pm$    0.016 &$\pm$     0.048&$\pm$    0.064 &$\pm$     0.133    \\
   9  &  -22.336    &    -0.023     &  -22.346    &     0.010     &  -22.450    &    -0.026     &  -22.260    &     0.580     &  -22.005    &     0.074     &  -21.918    &     0.242    \\
  &$\pm$    0.014 &$\pm$     0.032&$\pm$    0.006 &$\pm$     0.026&$\pm$    0.017 &$\pm$     0.037&$\pm$    0.037 &$\pm$     0.106&$\pm$    0.007 &$\pm$     0.031&$\pm$    0.047 &$\pm$     0.132    \\
  10  &  -22.409    &    -0.032     &  -22.428    &    -0.004     &  -22.557    &    -0.070     &  -22.562    &     0.109     &  -22.019    &     0.032     &  -21.980    &     0.078    \\
  &$\pm$    0.016 &$\pm$     0.024&$\pm$    0.007 &$\pm$     0.017&$\pm$    0.019 &$\pm$     0.027&$\pm$    0.020 &$\pm$     0.050&$\pm$    0.009 &$\pm$     0.020&$\pm$    0.019 &$\pm$     0.049    \\
  11  &  -22.358    &     0.057     &  -22.332    &     0.042     &  -22.553    &     0.005     &  -22.507    &     0.120     &  -21.969    &     0.078     &  -21.882    &     0.162    \\
  &$\pm$    0.015 &$\pm$     0.020&$\pm$    0.006 &$\pm$     0.013&$\pm$    0.018 &$\pm$     0.023&$\pm$    0.012 &$\pm$     0.031&$\pm$    0.008 &$\pm$     0.013&$\pm$    0.010 &$\pm$     0.028    \\
  12  &  -22.348    &    -0.091     &  -22.389    &    -0.079     &  -22.536    &    -0.144     &  -22.646    &    -0.034     &  -22.049    &    -0.012     &  -22.098    &    -0.110    \\
  &$\pm$    0.025 &$\pm$     0.038&$\pm$    0.010 &$\pm$     0.029&$\pm$    0.028 &$\pm$     0.041&$\pm$    0.047 &$\pm$     0.108&$\pm$    0.013 &$\pm$     0.033&$\pm$    0.052 &$\pm$     0.121    \\
  13  &  -22.469    &     0.123     &  -22.419    &     0.110     &  -22.531    &    -0.072     &  -22.412    &     0.441     &  -22.031    &     0.044     &  -21.936    &     0.224    \\
  &$\pm$    0.032 &$\pm$     0.050&$\pm$    0.012 &$\pm$     0.041&$\pm$    0.039 &$\pm$     0.059&$\pm$    0.063 &$\pm$     0.167&$\pm$    0.019 &$\pm$     0.057&$\pm$    0.077 &$\pm$     0.207    \\
  14  &  -22.401    &     0.029     &  -22.386    &     0.011     &  -22.551    &    -0.044     &  -22.661    &    -0.169     &  -21.939    &     0.080     &  -21.883    &     0.093    \\
  &$\pm$    0.019 &$\pm$     0.032&$\pm$    0.006 &$\pm$     0.025&$\pm$    0.022 &$\pm$     0.035&$\pm$    0.054 &$\pm$     0.107&$\pm$    0.010 &$\pm$     0.032&$\pm$    0.064 &$\pm$     0.129    \\
  15  &  -22.538    &     0.215     &  -22.436    &     0.176     &  -22.671    &     0.077     &  -22.502    &     0.272     &  -22.038    &     0.077     &  -21.966    &     0.133    \\
  &$\pm$    0.021 &$\pm$     0.032&$\pm$    0.008 &$\pm$     0.024&$\pm$    0.022 &$\pm$     0.033&$\pm$    0.032 &$\pm$     0.070&$\pm$    0.009 &$\pm$     0.022&$\pm$    0.030 &$\pm$     0.062    \\
  16  &  -22.489    &     0.067     &  -22.459    &     0.058     &  -22.554    &    -0.113     &  -22.565    &     0.138     &  -22.098    &     0.070     &  -22.000    &     0.203    \\
  &$\pm$    0.026 &$\pm$     0.041&$\pm$    0.008 &$\pm$     0.029&$\pm$    0.031 &$\pm$     0.050&$\pm$    0.044 &$\pm$     0.106&$\pm$    0.011 &$\pm$     0.032&$\pm$    0.041 &$\pm$     0.101    \\
  \hline
\end{tabular}
\end{table*}

  The final photometric catalog in the MKO system  has been obtained  by imposing these coefficients
and the analogous equations to the instrumental magnitudes of all detected objects.
\end{appendix}
  \end{document}